\documentclass[twocolumn,aps,prd,tighten,floatfix,amssymb,amsmath,nofootinbib,superscriptaddress,10pt]{revtex4-1}

\usepackage{amssymb,bm}
\usepackage{graphicx}
\usepackage[usenames]{color}
\usepackage{url}
\usepackage{enumerate}
\usepackage{color}
\usepackage{xcolor}

\newcommand{\sign}{{\rm sgn}}

\begin{document}

\title{Gradient sub-grid-scale model for relativistic MHD Large Eddy Simulations}

\author{Federico Carrasco}
\affiliation{Departament  de  F\'{\i}sica, Universitat  de  les  Illes  Balears  and  Institut  d'Estudis Espacials  de  Catalunya,  Palma  de  Mallorca, Baleares  E-07122,  Spain}
\affiliation{Institut d'Aplicacions Computacionals de Codi Comunitari (IAC3),  Universitat  de  les  Illes  Balears,  Palma  de  Mallorca,  Baleares  E-07122,  Spain}
%\affiliation{Max Planck Institute for Gravitational Physics (Albert Einstein Institute), \\
%Am M\¨{u}hlenberg 1, 14476 Potsdam, Germany}
\affiliation{Max Planck Institute for Gravitational Physics, % (Albert Einstein Institute),
Am M\"{u}hlenberg 1, 14476 Potsdam, Germany}

\author{Daniele Vigan\`o}
\affiliation{Departament  de  F\'{\i}sica, Universitat  de  les  Illes  Balears  and  Institut  d'Estudis Espacials  de  Catalunya,  Palma  de  Mallorca, Baleares  E-07122,  Spain}
\affiliation{Institut d'Aplicacions Computacionals de Codi Comunitari (IAC3),  Universitat  de  les  Illes  Balears,  Palma  de  Mallorca,  Baleares  E-07122,  Spain}

\author{Carlos Palenzuela}
\affiliation{Departament  de  F\'{\i}sica, Universitat  de  les  Illes  Balears  and  Institut  d'Estudis Espacials  de  Catalunya,  Palma  de  Mallorca, Baleares  E-07122,  Spain}
\affiliation{Institut d'Aplicacions Computacionals de Codi Comunitari (IAC3),  Universitat  de  les  Illes  Balears,  Palma  de  Mallorca,  Baleares  E-07122,  Spain}

%\affiliation{${^1}$ Departament  de  F\'{\i}sica, Universitat  de  les  Illes  Balears  and  Institut  d'Estudis Espacials  de  Catalunya,  Palma  de  Mallorca, Baleares  E-07122,  Spain}
%\affiliation{$^{2}$ Institut d'Aplicacions Computacionals de Codi Comunitari (IAC3),  Universitat  de  les  Illes  Balears,  Palma  de  Mallorca,  Baleares  E-07122,  Spain}

\date{January 2019}

\begin{abstract}
MHD turbulence is likely to play an important role in several astrophysical scenarios, where the magnetic Reynolds is very large. Numerically, these cases can be studied efficiently by means of Large Eddy Simulations, in which the computational resources are used to evolve the system only up to a finite grid size. The resolution is not fine enough to capture all the relevant small-scale physics at play, which is instead effectively modeled by a set of additional terms in the evolution equations, dubbed as sub-grid-scale model. Here we extend such approach, commonly used in non-relativistic/non-magnetic/incompressible fluid dynamics, to any general set of equation written in conservative form. We apply the so-called gradient model, giving recipes for these general balance-law systems, including the relevant case in which a non-trivial inversion of conserved to primitive fields is needed.
In particular, we focus on the relativistic compressible ideal MHD scenario, by providing for the first time and for any equation of state, all the additional non-trivial sub-grid-scale terms. 
As an application, we consider box simulations of the relativistic Kelvin-Helmholtz instability (KHI), which is also the first mechanism responsible for the magnetic field amplification in binary neutron star mergers and cannot be captured by the finest-grid and longest simulations available (currently and in the near-future). We numerically assess the performance of our model, by comparing it to the residuals coming from the filtering of high-resolution simulations. We find that the model can fit very well the residuals coming from filtering simulations with a resolution a few times higher.
The application shown here explicitly considers the Minkovski metric, but it can be directly extended to general relativity, thus settling the basis to implement for the first time the gradient sub-grid model in a GRMHD binary merger LES. Our results suggest that this approach will be potentially able to unveil much better the small-scale dynamics achievable of full GRMHD simulations, or equivalently, to obtain the same results but saving a considerable amount of computational time.
\end{abstract}

\maketitle

\label{firstpage}

\section{Introduction}

The first long-awaited simultaneous detection of gravitational waves (GW) and electromagnetic radiation from a binary neutron star (BNS) merger \cite{abbott17a} contains a wealth of data, including multi-wavelength electromagnetic radiation ranging from radio to gamma \cite{abbott17b}. 
The existing models of BNS mergers allow to compare the observed GW signal with the theoretical waveform, thus inferring the chirp mass, and, secondarily, constraining the mass ratio and the deformability of nuclear matter.
This analysis, together with the associated short gamma-ray burst  and the kilonova emission~\cite{metzger17}, served to constrain some unanswered questions, regarding for instance the equation of state of nuclear matter \cite{radice18}, the maximum mass supported by a neutron star \cite{margalit17,ruiz18} and its radius \cite{bauswein17}, the amount of ejecta \cite{perego17,cote18} and the contribution of BNS mergers to heavy elements production via r-processes \cite{abbott17c}.

Given the intrinsic difficulties of numerical general relativistic magneto-hydrodynamics (GRMHD) simulations, a few works have consistently included the full set of combined Einstein equations and MHD equations to solve such scenarios. Although the main features of the dynamics are quite clear from these simulations, some interesting and more subtle questions regarding the role of magnetic fields and the process they undergo after the merger are still unclear. In particular, during the merger, the Kevin-Helmholtz instability (KHI) is triggered at the shear layer between the colliding stars, and induces a fast growth of any seed magnetic field~\cite{price06,2008PhRvL.100s1101A,giacomazzo15,kiuchi15}. The instability develops faster for small-scale perturbation, with a cut-off wavelength of the order of the shear layer thickness (likely being $\sim $ meter or less). Additionally, other two mechanisms related to the remnant's rotation are supposed to play an important role at longer timescales: the magnetic winding, which creates and amplifies toroidal magnetic field  starting from a radial component, and the magneto-rotational instability (MRI)~\cite{balbus91}. 

A fundamental component of most theoretical models of binary NS mergers involved the formation of a jet perpendicular to the orbital plane, which is expected to be produced due to the accretion of the disk onto a central black hole. This jet is supposed to power the short gamma ray burst \cite{lazzati17,ciolfi18}, and it is often thought to require or be favoured by the presence of a strong, large-scale (dipolar) magnetic field. The creation of the jet is not trivial, with some simulations succeeding in generating it when starting with a strong magnetic field \cite{ruiz16}, but most of them not yet able to see it (see for instance \cite{kawamura16}). In any case, the growth and creation of a large scale magnetic field has not yet been observed, not even in simulations with the highest numerical resolutions of O(10m)~\cite{kiuchi15}. Even in that case,  they are not yet able to fully capture the KHI, whose smallest and fastest-growing scales are of the order of the discontinuity layer (probably a few meters or smaller). 
%The differences in timescales are the main reason behind the difficulties in simulating all processes at play in a consistent way. Since the fastest growing mode of MRI has a wavelength proportional to $B$, the higher it is, the easier is to capture it with at least 10 points (a common criterion to guarantee a proper capture). This feature explains why several studies of MRI in the post-merger disks start with a large-scale magnetic field of the order $10^{15}-10^{16} \gs$, justifying this value on simulations that reach it locally, for small-scale magnetic structures.

Limitations are being slowly overcome thanks to the growing computational resources. However, the full capture of all scales by means of direct numerical simulation (DNS) is still relatively far out of reach even for the most advanced codes and infrastructures. Thus, it is well worth to study alternative approaches that try to simulate the physical processes at play with a much lower computational cost.
There have been several attempts to effectively model the dynamics occurring in the small scales. For instance, Ref.~\cite{shibata17a,shibata17b} employs the viscous hydrodynamic equations to model the differentially rotating remnants of binary neutron stars mergers. 
Other works concern dynamo mechanisms, in which
the magnetic field growth can be modeled by a simple algebraic modification of the induction equation~\cite{palenzuela15,giacomazzo15}.

An alternative approach to the problem comes from computational fluid dynamics and it can be summarized in the following statements. First, any numerical simulation can be seen as a filtered version of the evolution equations. Secondly, the filtering of any non-linear combination of the evolved fields (e.g., in the fluxes of the MHD equations) implies the appearance of residual terms in the evolution equations, corresponding to the unresolved dynamics in the small scales. In this approach, commonly known as explicit Large Eddy Simulations (LES), relies first on this separation of the scales of the solution (i.e., resolved and unresolved), combined with an explicit sub-grid scale (SGS) model to account for the dynamics occurring at the unresolved small scales~\cite{mason1994large,lesieur1996new,sagaut2006large,bersellimathematics}.
Although LES are commonly used in many engineering applications, including  combustion, acoustics, and simulations of the atmospheric boundary layer and other fluids, its extension to astrophysical magnetized plasmas is less spread \cite{miesch15}. Its application to strongly self-gravitating fluids is even more limited. To our knowledge, only Ref.~\cite{radice17} has so far explicitly developed explicit LES for the relativistic hydrodynamic equations with a covariant form of the viscous SGS model, proposed by Smagorinsky in the sixties \cite{smagorinsky63} and commonly used in other contexts.
Since the Smagorinsky model is by definition purely dissipative, it easily allows numerical stability, but it cannot capture any possible backscatter towards larger scales. This is one of the reason why more elaborated models have been considered in the non-relativistic case.

In this paper we propose a more general formulation of the so-called gradient model \cite{leonard75,muller02a}, that we already assessed and implemented in the non-relativistic case \cite{paper1}. It allows to capture part of the unresolved small-scale dynamics, regardless of the specificity of the problem. The main advantage of this model is that it does not rely on any physical assumption in the functional form of these residual terms.
First we have to extend the LES formalism to generic conservative evolution equations like the relativistic MHD ones. This system has some particularities; first, the evolution equations of the evolved fields depend on some other quantities, usually called primitive fields. Second, the relation between the conserved and the primitive fields is explicit and analytical, but the inverse is usually not. 
After writing the LES for generalized conservation equations, we will extend the gradient model to this system.

%{\color{blue}
Let us emphasize again our aim: assuming that the relativistic MHD equations can be written as an evolution system by using the 3+1 decomposition, how can we efficiently mimic the effect of the small-scale dynamics, which are missed due to the finite numerical resolution? The goal of this work is, hence, to extend the use of a well established numerical technique (i.e. LES and a particular SGS model) to relativistic MHD. 
It is worth to discuss here some possible limitations of our approach. First, our proposed SGS model for relativistic MHD is not gauge invariant. The construction of the SGS model (and, more generally, any LES) begins from the filtering operation, which is to be associated in our context with the discretization of the equations at the 3+1 level. Thus, the filtering operator is not invariant but depends on the foliation, and therefore, the SGS tensors arising from such filtering are not gauge invariant neither. That means that different gauges will have different averages and, as a consequence, different SFS tensors. Since the SGS terms are intended to effectively capture part of the missing small scale dynamics (contained in the SFS tensors), it is quite natural to expect a gauge dependent SGS model, adapted instead to the particular discretization of the system at the 3+1 level. 
It is worth to reiterate that the gradient model does not modify the continuous limit, nor the principal part of the evolution system. In this sense, the inclusion of the SGS terms is analogous to the numerical reconstruction methods commonly used in the HRSC schemes, with the important difference that a careful analysis of the equations needs to be done in order to derive the functional forms of the SGS terms. Once the latter is done, it can be applied to any system of hyperbolic PDEs after the 3+1 decomposition.
Another important point that we want to stress is that the LES non-relativistic MHD model has the mathematical properties (conservation of total energy, magnetic helicity, approximate conservation of the cross helicity) expected of a model derived from the MHD equations by an averaging operation \cite{LABOVSKY2011516}. Similar extensions would be expected to hold in the relativistic case.
Finally, probably the strongest limitation which could affect the effectiveness of SGS model is the presence of strong shock in the simulations. Although the High-Resolution-Shock-Capturing methods that we use can deal with them by reducing the order of accuracy, this introduces an additional dissipation in the numerical scheme that overcomes the effects of the SGS model, as it is discussed for instance in~\cite{doi:10.1146/annurev-fluid-122109-160718}.
%}

In \S~\ref{sec:filtering}, we summarize the formalism of the filtering process in LES, very common in a plethora of hydro-dynamical simulations related to different fields, but less known in relativistic hydrodynamics. \S~\ref{sec:gradient} summarizes the basics of the gradient SGS model. In \S~\ref{sec:formulation}, we extend the LES formalism and the SGS gradient model to a general system of conservation laws, applied to non-relativistic and relativistic compressible ideal MHD in \S~\ref{sec:applications}. We validate the proposed model in a set of box simulations of the relativistic KHI in
\S~\ref{sec:simulations}, by comparing the SGS model to the residuals coming from filtering high-resolution simulations in \S~\ref{sec:apriori}. Finally, we summarize our results in 
\S~\ref{sec:conclusions}.

%%%%%%%%%%%%%%%%%%%%%%%%%%%%%%%%%%%%%%%%%%%%%%%%%%%%%
\section{Large Eddy Simulations and sub-filter-scale residuals}\label{sec:filtering}
%%%%%%%%%%%%%%%%%%%%%%%%%%%%%%%%%%%%%%%%%%%%%%%%%%%%%

From a formal point of view, the effect of finite resolution in a numerical simulation is equivalent to a low-pass spatial filter, with a filter size of the order of the grid cell. Applied to a field $f({\bf x},t)$, the filtering operation separates the resolved part from the sub-filter scale (SFS) residuals: \begin{equation}
f({\bf x},t)= \overline{f}({\bf x},t) + f'({\bf x},t)~.
\end{equation}
Indicating the filter kernel with $G$, the filtering operator over a field $f$ can be written generically as
\begin{equation}
 {\overline f}({\bf x},t) =  \int_{-\infty}^{\infty} G({\bf x}-{\bf x'}) f({\bf x'},t) d^3 x' ~.
\end{equation}
Large-Eddie-Simulations consist on applying the filter to the evolution equations under consideration and write them down as a function only of the resolved fields. 

Let us consider a generic evolution system written as
\begin{equation}
\partial_t U^a + \partial_k F^{ka} (U) = 0~, \label{eq:conservation_law}
\end{equation}
where $U^a$ is a set of evolved fields and $F^{ka}(U)$ is the flux along the direction $k$ of the field $U^a$. The equation for the filtered fields can be obtained by applying the filtering operator to the equations. Since the filtering operator commutes with both spatial and time derivatives, we have 
\begin{equation}
\partial_t \overline{U}^a + \partial_k F^{ka} (\overline{U}) = \partial_k \overline{\tau}^{ka}
\end{equation}
%where the indexes ``$a,b,e$'' represents the different fields of the set of evolved quantities and ``$i,j,k$'' represents the spatial directions.
where we have defined the SFS tensors related to the fluxes as
\begin{equation}\label{residual-F}
\overline{\tau}^{ka}_{F}:=  F^{ka} (\overline{U}) -  \overline{F^{ka}(U)} ~.
\end{equation}

In order to see the effect of this scale-separation in a simple application, let us consider the well-known Burgers equation, a longstanding field of research in LES \cite{love80}. In the one-dimensional case, given the velocity field $u$, the equation reads: 
\begin{equation}\label{eq:burgers}
\partial_t u + \frac{1}{2} \partial_x u^2 = 0~.
\end{equation}
The equation for the filtered fields can be obtained by applying the filtering operator to the equations, namely
\begin{equation}\label{eq:burgers_filter}
\partial_t \overline{u} + \frac{1}{2}  \partial_x  \overline{u}^2 = \partial_x \overline{\tau}~~~,~~~\overline{\tau} \equiv \overline{u}^2 - \overline{u^2} ~.
\end{equation}
Notice that the filtered equation is equivalent to the original one except by a term proportional to  $\overline{\tau}$ which represents the SFS residuals, that is, the loss of small-scale information due to the filtering process related to the non-linear terms. If in the region over which we average $u$ has a definite sign (which is usually the case except close to the shock), then $\overline{u}^2 < \overline{u^2} $ and $\overline{\tau}$ is negative-definite, so that $-\overline{\tau}$ represents the kinetic energy hidden in the SFS.

Regardless of the specific system considered, these new terms always appear due to the non-commutativity of the filtering operator with the non-linear terms of the equations. In a LES, which effectively only evolves $\overline{u}$ and cannot simulate the unresolved scales, the explicit expression of $\overline{\tau}$ is a-priori unknown, and needs to be written as a function of the filtered fields in order to close the system.
This implies in practice to approximate the SFS tensors with a sub-grid scale (SGS) model.

There are several ways of doing that, based mostly either on physical arguments or on an expected self-similarity of the solution. For instance, a popular, historical approach to fluid dynamics~\cite{smagorinsky63} is the setup of an artificial viscosity $\nu$, namely
\begin{equation}\label{eq:burgers_viscosity}
\overline{\tau} = \overline{u}^2 - \overline{u^2} \approx  \nu \partial_{x} {\overline u} ~,
\end{equation}
The viscosity parameter is often taken to be proportional to $\Delta^2|\partial_{x} {\overline u}|$, due to dimensional reasons~\cite{smagorinsky63} and in order to ensure that this term vanishes in the continuum limit  $\Delta \rightarrow 0$, thus guaranteeing numerical convergence to the continuum (or DNS) solution. The numerical value of the proportionality coefficient can be fixed by hand or estimated by means of dynamical procedures which assume self-similarity (thus similar to a multiscale/multivariational approach)~\cite{germano91}. Regardless of the value of $\nu$, this functional form of $\tau$ is equivalent to the viscous Burger's equation.

However, in many cases the physics involved does not consist only in dissipation, but can involve, for instance, inverse cascade and scale-dependent transfers of energy between the fluid kinetic energy and the magnetic one. Therefore, in those cases any result coming from the LES might be biased by the a-priori choice of a given SGS model.
%
%There are two approaches to address this problem: either the residuals are ignored, leaving only the intrinsic dissipation to act (i.e., the so-called implicit LES), or they are modeled by an additional sub-grid scale (SGS) tensor to close the equation. 
%The functional form \CP{and the numerical techniques?} for the SGS tensors is free and many have been proposed in the literature. 
Hereafter, we will focus on the so-called gradient model, which has already been shown to capture the main features of the turbulent dynamics in our previous study for non-relativistic MHD turbulence box simulations \cite{paper1}. 

%\subsection{Gradient SGS model}
\section{The SGS gradient model}
\label{sec:gradient}

Although many of the SFS terms are modeled based on some physical properties, it is possible to compute them relying only on mathematical arguments by considering the analytical Taylor expansion of the SFS terms\cite{leonard75,muller02a}, which rely on the properties of the filtering operator $G$. 

An homogeneous isotropic low-pass filter is independent on the direction (i.e., $G({\bf x}-{\bf x'})=G(|{\bf x}-{\bf x'}|)$) and only smooths out fluctuations on length scales smaller than the filter size, leaving unchanged the variations of the solution at larger length scales. In addition, the filter operator is linear and commutes with spatial derivatives. Generically, it can be written for any dimension $D$ as
\begin{eqnarray}
G(|{\bf x}-{\bf x'}|) = \prod_{i=1}^{D} G_i(|x_i-x'_i|) ~,
\label{eq:kernel_dim}
\end{eqnarray}
where $G_i(|x_i-x'_i|)$ is just the one-dimensional kernel function.

The simplest low-pass filter is the mean value in a cubic domain with size $\Delta_f$ in each Cartesian direction $\{x_i\}$, described by the normalized kernel
\begin{eqnarray}
G_i(|x_i-x'_i|) = 
\begin{cases}
1/\Delta_f ~~~~ {\rm if} ~~|x_i-x'_i| \le \Delta_f/2\\
0 ~~~~~~~~~ {\rm otherwise}
\end{cases}
\label{eq:kernel_space_box}
\end{eqnarray}
Despite the appealing simplicity of the box filter, which makes it very useful to perform numerical calculations, we will see below that it is not suitable for analytical calculations involving its derivatives, since they are not continuous. Therefore, at a formal level, it is more practical to introduce the normalized Gaussian kernel, which in the space domain can be written as
\begin{eqnarray}
G_i(|x_i-x'_i|) =  \left( \frac{1}{4 \pi \xi } \right)^{1/2}
\exp\left(\frac{-|x_i-x'_i|^2}{4 \xi}\right) ~,
\label{eq:kernel_space_gaussian}
\end{eqnarray}
where $\xi$ defines the effective filtering width. Besides having the same zeroth and first moments, Gaussian and box filters have the same second moment if we set $\xi=\Delta_f^2/24$.
The filter is a useful mathematical tool to analyze not only the different scales on the solution, but also the structure of the evolution equations, allowing us to distinguish between the resolved or filtered fields and the unresolved ones on the SFS.

The main hypothesis is that the discretization of the equations over a finite-size grid is approximated by a filtering operator acting on the equations, with a Gaussian kernel equivalent to a box filter with width $\Delta_f$, eq.~(\ref{eq:kernel_space_gaussian}). Its $D$-dimensional Fourier transformed function is
\begin{eqnarray}
{\hat G}({\bf k}) = \exp(- \xi {\bf k}^2) ~,
\label{eq:kernel_fourier}
\end{eqnarray}
where $\bf k$ is the wavenumber.
The main idea is to compute an approximation of the inverse filtering operator based on gradient expansion of the filter kernel $G$, that is, an approximation of the inverse Fourier transform of $1/{\hat G}$. The first step is to perform a Taylor expansion of the transformed function and its inverse in terms of the filter scale, that is,
\begin{eqnarray}\label{eq:kernel_expansion}
{\hat G}({\bf k}) &=&  \sum_{n=0}^{\infty} \frac{(-1)^{n}}{n!} 
\left(\xi {\bf k}^2 \right)^{n} ~, \\
\frac{1}{{\hat G}({\bf k})} &=&  \sum_{n=0}^{\infty} \frac{1}{n!} 
\left(\xi {\bf k}^2 \right)^{n} ~.
\end{eqnarray}
Considering the expansions to the fields ${\hat f}$ and ${\hat {\overline f}}$, the application of the inverse Fourier transformation yields to an infinite series representation of the filter operator and its inverse in terms of gradient operators acting on the fields, namely\cite{vlaykov16}
\begin{eqnarray}
{\overline f} \equiv G * f &=&  \sum_{n=0}^{\infty}
\frac{1}{n!} 
\left(\xi \nabla^2 \right)^{n} f  ~, \\
f \equiv G^{-1} * {\overline f} &=&  \sum_{n=0}^{\infty}
\frac{(-1)^{n}}{n!} 
\left(\xi \nabla^2 \right)^{n} {\overline f} ~~.
\label{eq:gradient_operator}
\end{eqnarray}
These expressions are absolutely convergent and formally accurate at all orders, since the Gaussian kernel is infinitely differentiable and with unbound support. In fact, it was found that these series converges for all canonical filters.

The main idea is to filter the evolution system by means of a spatial filtering with Gaussian kernel (e.g.~\cite{grete16,grete17}). Then, there appear unknown SFS tensors, which can be computed explicitly by using the Taylor expansion Eqn. \eqref{eq:gradient_operator}.
Therefore, we have obtained the following useful relations to first order in $\xi$:
\begin{eqnarray*}
	\overline{fg} &\backsimeq&  {\overline f} \, {\overline g} + 2\, \xi\, \nabla {\overline f} \cdot \nabla {\overline g}~, \\
	\overline{fgh} &\backsimeq&  {\overline f} \, {\overline g} \, {\overline h} + 2\, \xi\, \left(   {\overline h} \, \nabla {\overline f} \cdot \nabla {\overline g} 
	+ {\overline g} \, \nabla {\overline f} \cdot \nabla{\overline h} + {\overline f} \, \nabla {\overline g} \cdot \nabla {\overline h} \right)~, \\
	\overline{f(g)} &\backsimeq&  f ({\overline g}) + \xi \left(  \nabla^2 f({\overline g}) - \frac{df}{d{\overline g}} \nabla^2 {\overline g} \right) \\
	&\backsimeq&  f ({\overline g}) + \xi \, \nabla \frac{df}{d{\overline g}} \cdot \nabla {\overline g}~,
\end{eqnarray*}
where ``$\backsimeq$'' means (here and thereafter) ``accurate up to $\mathcal{O}(\xi^2)$ terms'' and $\nabla () \cdot \nabla () $ denotes contractions of spatial derivatives.
Also, note that by $\frac{df}{d{\overline g}}$ we really mean $\frac{df}{dg}({\overline g})$.

Going back to our previous Burger's equation example, it is easy to model the SFS terms appearing in Eq.~(\ref{eq:burgers_filter}) using the gradient expansion above, namely
\begin{equation}\label{eq:burgers_filter2}
\overline{\tau} \equiv \overline{u}^2 - \overline{u^2} \backsimeq - 2\, \xi\, \partial_x {\overline u} \, \partial_x {\overline u}~.
\end{equation}
In principle, corrections to these expression are expected due to the form factor (due to the kernel shape) and the contribution from higher orders.
In other words, this approximated expression can help as long as we can capture most of the dynamics with our LES, leaving to the SGS the task of mimicking the small-scale contributions.
Notice that the SGS terms scale with $\xi\propto \Delta_f^2$, thus ensuring the numerical convergence (i.e., it vanishes in the continuous limit $\Delta_f\rightarrow 0$).

Regardless on these quite obvious caveats, notice that this prescription is very different from the viscous one given by eqn. \eqref{eq:burgers_viscosity}: one arises from physical considerations, while the other just from mathematical ones. The first one is very useful in scenarios where the physics involved is well known and present universal behaviour, like in pure hydro-dynamical, non-relativistic cases. However, in the case of MHD, the non-linear interplay between kinetic and magnetic energy is rich of different physical mechanisms (dynamo effect, dissipation, helicity transfer...), at different scales. In these cases, it is more useful to have a SGS model which relies only on the mathematical features (basically, the gradients) of the involved field.

%Remember that we need to compute the following SFS terms, eqs.~(\ref{eq:tau_kin_final})-(\ref{eq:tau_p_final}) and (\ref{eq:sigma_press}), namely
%\begin{eqnarray}
%&&  - \overline{\tau}_{\rm kin}^{ki} =  \overline{\rho v^k v^i} - \overline{\rho}\widetilde{v}^k \widetilde{v}^i \label{eq:tau_kin_final2}\\
%&&  - \overline{\tau}_{\rm mag}^{ki} =  \overline{B^k B^i} - \overline{B}^k \overline{B}^i \label{eq:tau_mag_final2} \\
%&&  - \overline{\tau}_{\rm ind}^{ki} =  (\overline{v^k B^i} - \overline{v^i B^k}) -  (\widetilde{v}^k \overline{B}^i - \widetilde{v}^i \overline{B}^k) \label{eq:tau_ind_final2} \\
% &&   - \overline{\tau}^k_{\rm enth} = \overline{h v^k} - [ \overline{\rho} (1+\widetilde{e}) + p(\overline{\rho},\widetilde{e})]\widetilde{v}^k \label{eq:tau_enthalphy_final2} \\
% && - \overline{\tau}_{\rm p}^{ki} =   [\overline{p} - p(\overline{\rho},\widetilde{\epsilon})]\delta^{ki} \\
% && - \overline{\Sigma}_{\rm pres} =  \overline{v^j \partial_j p} - \widetilde{v}^j\partial_j p(\overline{\rho},\widetilde{e})
%\label{eq:sigma_press_final2} 
%\end{eqnarray}

%%%%%%%%%%%%%%%%%%%%%%%%%%%%%%%%%%%%%%%%%%%%

%%%%%%%%%%%%%%%%%%%%%%%%%%%%%%%%%%%%%%%%%%%%%%%%%%%%%
\section{Gradient model for general system of conservation laws} \label{sec:formulation}
%%%%%%%%%%%%%%%%%%%%%%%%%%%%%%%%%%%%%%%%%%%%%%%%%%%%

The gradient model described before has been applied mainly to the non-relativistic HD or MHD equations, often in their incompressible version. 
In order to generalize it to the relativistic case, we tackle the problem from a broader perspective by considering a general system of conservation laws, 
\begin{equation}
\partial_t C^a + \partial_k F^{ka} (P) = 0~, \label{eq:conservation_law}
\end{equation}
where $C^a$ is a set of {\em conserved} evolved variables and $F^{ka}(P)$ is the flux along the direction $k$ of the field $C^a$, which can be expressed in terms of the {\em primitive} fields $P^a$. The transformations between conserved and primitive fields can be expressed formally as
\begin{equation*}
	C^a = f^a (P) ~,~~~
	P^a :=  (f^{-1})^{a}(C) \equiv g^{a}(C)~.
\end{equation*}
Notice that, although the relation $f^a$ between conserved and primitive fields is always known explicitly, the inverse function $g^a$ is not analytical in the relativistic case, where it usually needs to be solved numerically.

Conservative-scheme simulations are then equivalent to a filtered version of the evolution equations, such that the values of the filtered conserved variables $\overline{C}^a$ are numerically known. 
The filtering operation, implicitly contained in a LES, allows us to evaluate $\widetilde{P}^a :=  g^a (\overline{C})$ but not   $\overline{P}^a \equiv \overline{g^{a}(C)}$, which is what really appears in the filtered version of the equations. 
This distinction suggests a different definition of the SFS tensor in the filtered equations, which
can be obtained again by applying the filter to eqs. \eqref{eq:conservation_law}, namely
\begin{equation}
\partial_t \overline{C}^a + \partial_k F^{ka} (\widetilde{P}) = \partial_k \overline{\tau}^{ka}~,
\end{equation}
where hereafter the indexes ``$a,b,e$'' represent the different fields of the set of primitive and conserved quantities, while ``$i,j,k$'' represent the spatial directions. Here we have defined the SFS tensors related to the fluxes as
\begin{equation}\label{residual-F}
\overline{\tau}^{ka}_F :=  F^{ka} (\widetilde{P}) -  \overline{F^{ka}(P)} ~~.
\end{equation}
that is, in terms of $\widetilde{P}^a$ instead of $\overline{P}^a$, which are the fields that we can compute from the evolved $\overline{C}^a$.
Notice that these SFS tensors for the fluxes are not special, and similar ones can be calculated for any function. In particular, such residuals can be computed for the primitive fields $P^a$ (or for any other non-conserved field), namely,
\begin{equation}
\overline{\tau}^{a}_P :=  g^a(\overline{C}) -  \overline{g^a(C)} ~. \label{eq:sfs_auxiliary_field}
\end{equation}
%where the sub-script to the $\overline{\tau}$ indicates the field to which the SFS residuals are referred, while that super-script indicates the specific component of the field.
where the labels on $\overline{\tau}$ indicates to which field (and specific component of the field) the SFS residuals are referred.

%{\color{blue} Here  $\widetilde{P}^a$ are the ``filtered primitive fields'' that we can compute from the filtered conservative ones through,
%\begin{equation*}
%\widetilde{P}^a :=  (f^{-1})^{a}(\overline{C}) \equiv g^{a}(\overline{C})~.
%\end{equation*}
%which differs from $\overline{P}^a \equiv \overline{g^{a}(C)}$.
%This inversion is not analytical in the relativistic case, where it usually needs to be solved numerically. 
%Nevertheless, in that case  the inversion from conserved to primitive fields was already needed for computing the fluxes in the conservative schemes.}
%\FC{quitaria esto de aca, para decirlo mas arriba donde ya lo estaban diciendo.. o al reves.}
% Notice that it differs from $\overline{P}^a \equiv \overline{g^{a}(C)}$, i.e.
%\begin{equation*}
%\overline{P}^a = \overline{g^a(C)} \backsimeq \widetilde{P}^a + \xi \left( \nabla^2 g^a(\overline{C}) - \frac{dg^a}{d\overline{C}^b} \nabla^2  \overline{C}^b \right)  
%\end{equation*}

In general, neither the SFS tensors associated to the primitive fields nor to the fluxes are known, and they can be important whenever a non-negligible part of the dynamics occurs in the small scales. 
This is the typical scenario in turbulence, where the fluctuations are relevant. 
The task is now to exploit the above gradient expansion of the filter kernel to provide a closed expression for $\overline{\tau}^{ka}_F$~\eqref{residual-F} in terms only of the filtered variables and its derivatives.
As we will see, this process essentially involves the filtering of composed functions; the Taylor expansion; and the inverse function theorem.

Let us start by expanding the expression
\begin{equation}\label{over_Ska}
 \overline{F^{ka}(P)}  \backsimeq F^{ka}(\overline{P}) + \xi \left( \nabla^2 F^{ka} (\overline{P}) - \frac{dF^{ka}}{d\overline{P}^{b}} \nabla^2  \overline{P}^b \right)   ~,
\end{equation}
where repeated index ``$a,b,..$'' denotes summation over the space of fields. Then, we can Taylor expand $F^{ka}(\overline{P})$ around $\widetilde{P}$, to first order in $\xi$, as 
 \begin{equation}
F^{ka}(\overline{P}) \backsimeq F^{ka}(\widetilde{P})  + \xi \, \frac{dF^{ka}}{d\widetilde{P}^b} \, \left(  \nabla^2 \widetilde{P}^b - \frac{d\widetilde{P}^b}{d\overline{C}^e} \nabla^2  \overline{C}^e \right)~, \label{taylor-multi}
\end{equation}
and by the inverse function theorem we can express (locally) the Jacobian of the inverse variable transformation as the inverse of the Jacobian. That is,
\begin{equation}
 \frac{d\widetilde{P}}{d\overline{C}} \equiv \left( \frac{dC}{dP}\right)^{-1} \bigg\rvert_{\widetilde{P}, \overline{C}} 
\end{equation}
meaning the matrix inversion of the Jacobian $\frac{dC^a}{dP^b}$, evaluated at the filtered variables (namely, either $\overline{C}^a$ or $\widetilde{P}^a$).

Finally, by noticing that we can interchange $\overline{P} \rightleftarrows \widetilde{P} $ (at first order in $\xi$) in the last term of \eqref{over_Ska}, and then combining it with \eqref{taylor-multi} we finally get the functional form of the SGS tensor $\tau^{ka}_F$, namely
\begin{equation}\label{tau-ka}
 \tau^{ka}_F = \xi \, \left( \frac{dF^{ka}}{d\widetilde{P}^b} \, \frac{d\widetilde{P}^b}{d\overline{C}^e} \, \nabla^2  \overline{C}^e - \nabla^2 F^{ka} (\widetilde{P})\right)  
\end{equation}
which approximates the SFS residuals of the fluxes  (i.e., $\tau^{ka}_F \backsimeq \overline{\tau}^{ka}_F$).
In most of the cases it is useful to re-express the SGS tensor $\tau^{ka}$ above in the following equivalent form
\begin{equation}\label{tau-alt} 
\tau^{ka}_F = -\xi \, \nabla  \frac{dF^{ka}}{d\overline{C}^b}  \cdot \nabla \overline{C}^b 
\end{equation}
Due to its simplicity we will use this relation in the following, although in some specific evolution systems other equivalent forms can be preferred\footnote{ For instance, another interesting equivalent expression is given by $\tau^{ka} = -\xi  \, \frac{d^2 F^{ka}}{d\overline{C}^b d\overline{C}^c} \, \nabla \overline{C}^b  \cdot \nabla \overline{C}^c $, which may be useful in situations where the fluxes can be expressed in terms of conserved quantities in a close form. 
	%\FC{..relacionar la propiedad de concavidad/convexidad de los flujos con los posibles efectos disipativos o cascada inversa de los SGS tensors resultantes. }  
}.

%%%%%%%%%%%%%%%%%%%%%%%%%%%%%%%%%%%%%%%%%%%%%%%%%%%%%%%%%%%%%%%%%%%%%
\section{Applications to compressible ideal MHD} \label{sec:applications}
%%%%%%%%%%%%%%%%%%%%%%%%%%%%%%%%%%%%%%%%%%%%%%%%%%%%%%%%%%%%%%%%%%%%%

We will check the validity of our approach by applying first to the non-relativistic MHD equations, for which the LES equations with the gradient model has been studied for decades and were derived in the compressible case for a generic equation of state~\cite{paper1}. After that, we will consider the unexplored MHD relativistic case.

\subsection{Non-relativistic case}

The set of primitives fields for the non-relativistic MHD is given by $P^a = \left\lbrace \rho, v^i , \epsilon, B^i  \right\rbrace $ (density, velocity, specific internal energy and magnetic field, respectively). 
The conserved ones are $C^a = \left\lbrace \rho, N^i , U , B^i \right\rbrace$ (density, momentum density, energy density and magnetic field), which can be written explicitly as a function of the primitive ones as:
\begin{equation}
 C^a =  f^a (P) =  \left\lbrace \rho, \rho v^i , \rho \epsilon + \rho \frac{v^2}{2} + \frac{B^2}{2}  , B^i \right\rbrace
\end{equation}
The evolution system is thus written as
\begin{eqnarray*}	
 && \partial_t \rho + \partial_k N^k = 0 ~, \\
 && \partial_t N^i + \partial_k T^{ki} = 0 ~,\\ 
 && \partial_t U + \partial_k S^{k} = 0 ~,\\
 && \partial_t B^i + \partial_k M^{ki} = 0 ~,
\end{eqnarray*}
where the fluxes read
\begin{eqnarray*}
 N^k &=& \rho v^k ~,\\
 T^{ki} &=& \rho v^i v^k - B^i B^k + \delta^{ki} \left[ p(\rho, \epsilon) + B^2 / 2 \right] ~,\\
 S^{k} &=& \left[ U + p(\rho, \epsilon) + B^2 / 2 \right]  \, v^k - (v \cdot B) B^k ~,\\
 M^{ki} &=& 2 \, v^{[i} B^{k]} ~, % B^i v^k - v^i B^k 
\end{eqnarray*}
with $2 v^{[i} B^{k]} := v^i B^k - B^i v^k $  denoting the usual anti-symmetrization operation. 
Notice that these evolution equations involve also the pressure $p(\rho,\epsilon)$, so an equation of state is required in order to close the system. The filtered version of the system, including the SFS terms in the right hand sides, is
\begin{eqnarray*}
  \partial_t \overline{\rho} + \partial_k N^k(\widetilde{P}) &=& \partial_k \overline{\tau}^{k}_{N} \\
  \partial_t \overline{N}^i + \partial_k T^{ki}(\widetilde{P})  &=& \partial_k \overline{\tau}^{ki}_{T} \\
  \partial_t \overline{U} + \partial_k S^{k}(\widetilde{P}) &=& \partial_k \overline{\tau}^{k}_{S} \\
  \partial_t \overline{B}^i + \partial_k M^{ki}(\widetilde{P}) &=& \partial_k \overline{\tau}^{ki}_{M} 
\end{eqnarray*}
In this problem the relation between conserved and primitives is easily invertible, allowing to express all the fluxes as explicit functions of the conserved quantities.
Thus, it is possible to compute $ \frac{d F^{ka}}{d\overline{C}^b} \equiv \left\lbrace \frac{d N^k}{d\overline{C}^b} , \frac{d T^{ki}}{d\overline{C}^b} , \frac{d S^{k}}{d\overline{C}^b} , \frac{d M^{ki}}{d\overline{C}^b}  \right\rbrace $,
and then use equation \eqref{tau-alt} to solve for
the SGS tensors $\tau_F^{ka}$. The former can be written explicitly as:
\begin{eqnarray*}
\frac{d N^k}{d\overline{C}^b} &\equiv& \left\lbrace  \frac{d N^k}{d \overline{\rho}}, \frac{d N^k}{d \overline{N}^j} , \frac{d N^k}{d \overline{U}}, \frac{d N^k}{d \overline{B}^j} \right\rbrace = \left\lbrace 0, \delta^{k}_{j} , 0, 0 \right\rbrace \\
\frac{d T^{ki}}{d\overline{C}^b}  &=&  \left\lbrace -\widetilde{v}^i \widetilde{v}^k , 2 \delta^{(i}_{j} \widetilde{v}^{k)} , 0, -2 \delta^{(i}_{j} \overline{B}^{k)} + \delta^{ki} \overline{B}_j \right\rbrace + \delta^{ki} \frac{d \widetilde{p}}{d\overline{C}^b} \\
\frac{d S^{k}}{d\overline{C}^b}  &=&  \left\lbrace -\frac{\widetilde{S}^k}{\overline{\rho}} , \frac{\widetilde{\Theta}}{\overline{\rho}} \delta^{k}_j - \frac{\overline{B}_j \overline{B}^k}{\overline{\rho}} , \widetilde{v}^k , \overline{B}_j \widetilde{v}^k - \widetilde{v}_j \overline{B}^k - (\widetilde{v}\cdot \overline{B}) \delta^{k}_j \right\rbrace \\
 && + \widetilde{v}^k \frac{d\widetilde{p}}{d\overline{C}^a} \\
 \frac{d M^{ki}}{d\overline{C}^b} &=&  \left\lbrace \frac{2}{\overline{\rho}} \, \overline{B}^{[i} \widetilde{v}^{k]}  , \frac{2}{\overline{\rho}} \delta^{[k}_{j} \overline{B}^{i]}  , 0, 2 \, \widetilde{v}^{[k}  \delta^{i]}_{j} \right\rbrace
\end{eqnarray*}
where $\widetilde{p} \equiv p(\overline{\rho}, \widetilde{\epsilon})$, $\widetilde{S}^k \equiv  S^k (\widetilde{P}) =   \widetilde{\Theta} \, \widetilde{v}^k - (\widetilde{v}\cdot \overline{B})\overline{B}^{k}$ and $\widetilde{\Theta} := \overline{U} + \widetilde{p} + \frac{\overline{B}^2}{2} $.
Also we will need the following derivative,
\begin{equation*}
 \frac{d \widetilde{p}}{d\overline{C}^b} =
% \left\lbrace \widetilde{p}_{\rho} - \frac{(\widetilde{\epsilon}-\widetilde{v}^2 /2)}{\overline{\rho}} \, \widetilde{p}_{\epsilon}, - \frac{\widetilde{v}_j}{\overline{\rho}} \, \widetilde{p}_{\epsilon}  ,  \frac{1}{\overline{\rho}} \, \widetilde{p}_{\epsilon} , - \frac{\overline{B}_j}{\overline{\rho}} \, \widetilde{p}_{\epsilon} \right\rbrace 
 \left\lbrace \frac{d\widetilde{p}}{d\overline{\rho}} - \frac{(\widetilde{\epsilon}-\widetilde{v}^2 /2)}{\overline{\rho}} \frac{d\widetilde{p}}{d\widetilde{e}}, - \frac{\widetilde{v}_j}{\overline{\rho}} \frac{d\widetilde{p}}{d\widetilde{\epsilon}}  ,  \frac{1}{\overline{\rho}}\frac{d\widetilde{p}}{d\widetilde{\epsilon}} , - \frac{\overline{B}_j}{\overline{\rho}} \frac{d\widetilde{p}}{d\widetilde{\epsilon}} \right\rbrace ~.
\end{equation*}
Finally, we obtain
\begin{eqnarray}
&& \tau^{k}_{N}  = - \xi \, \nabla \frac{d \overline{N}^k}{d\overline{C}^b} \cdot \nabla \overline{C}^b = 0~, \\
&& \tau^{ki}_{T} = - \xi \, \nabla \frac{d T^{ki}}{d\overline{C}^b} \cdot \nabla \overline{C}^b = \tau^{ki}_{\rm kin} - \tau^{ki}_{\rm mag} + \delta^{ki} \tau_{\rm pres}~, \\
&& \tau^{k}_{S}  = - \xi \, \nabla \frac{d S^{k}}{d\overline{C}^b} \cdot \nabla \overline{C}^b = \tau^{k}_{\rm ener} + \widetilde{v}^k \tau_{\rm pres}~, \\
&& \tau^{ki}_{M} = - \xi \, \nabla \frac{d M^{ki}}{d\overline{C}^b} \cdot \nabla \overline{C}^b = \tau^{ki}_{\rm ind}~,
\end{eqnarray}
where the SGS tensors have been splatted conveniently in order to connect with previous works, being:
\begin{eqnarray}
\tau^{ki}_{\rm kin} &=& - 2 \, \xi \, \overline{\rho} \, \nabla \widetilde{v}^{i} \cdot \nabla \widetilde{v}^{k}~, \\
\tau^{ki}_{\rm mag} &=& - 2 \, \xi  \, \nabla \overline{B}^{i} \cdot \nabla \overline{B}^{k} ~, \\
\tau_{\rm pres} &=& - \xi \, \nabla \frac{d\widetilde{p}}{d\overline{C}^b} \cdot \nabla \overline{C}^b + \frac{1}{2} \delta_{ls}\, \tau_{\rm mag}^{ls} \nonumber\\
&=& -\xi \, \left[ \nabla \frac{d\widetilde{p}}{d\widetilde{\rho}} \cdot \nabla \overline{\rho} + \nabla \frac{d\widetilde{p}}{d\widetilde{\epsilon}} \cdot \nabla \widetilde{\epsilon} - \frac{2}{\widetilde{\rho}}\,\frac{d\widetilde{p}}{d\widetilde{\epsilon}} \, \nabla \widetilde{\rho} \cdot \nabla \widetilde{\epsilon} \right.   \\
&& + \left. \nabla \overline{B}_{j} \cdot \nabla \overline{B}^{j} -  \frac{1}{\widetilde{\rho}} \, \frac{d\widetilde{p}}{d\widetilde{\epsilon}} \left( \overline{\rho} \, \nabla \widetilde{v}_{j} \cdot \nabla \widetilde{v}^{j}  + \nabla \overline{B}_{j} \cdot \nabla \overline{B}^{j} \right) \right]  \nonumber ~, \\
\tau^{k}_{\rm ener} &=& -2 \, \xi \, \left[ \nabla \widetilde{\Theta} \cdot \nabla \widetilde{v}^k + (\overline{B}^k \overline{B}_j \nabla \widetilde{v}^j - \widetilde{\Theta}  \nabla \widetilde{v}^k ) \cdot \nabla(\ln \widetilde{\rho})  \right. \nonumber \\
&& - \left. \overline{B}^k \nabla \overline{B}_j \cdot \nabla \widetilde{v}^j - \nabla (\widetilde{v}\cdot \overline{B}) \cdot \nabla \overline{B}^k \right]\\
 \tau^{ki}_{\rm ind} &=& - 4 \, \xi \, \left[ \nabla \widetilde{v}^{[k} \cdot \nabla \overline{B}^{i]}  + \overline{B}^{[i} \nabla \widetilde{v}^{k]} \cdot \nabla(\ln \overline{\rho}) \right] ~.
\end{eqnarray}
Notice that these results, obtained starting from a general formulation and applying it to the non-relativistic MHD case, agree with previous derivations calculated for this specific case  \cite{paper1}. The results here were derived for a general equation of state, which is reflected on the expression for $\tau_{\rm pres}$. In particular, the form of $\tau_{\rm pres}$ is greatly simplified for an ideal gas equation of state.

\subsection{Relativistic case}

Let us now consider the special relativistic MHD, for which a gradient SGS model has not been calculated so far. In this case, the set of primitives variables is given by $P^a = \left\lbrace \rho, v^i , \epsilon, B^i \right\rbrace $,
the conserved ones are $C^a = \left\lbrace D, S^i , U, B^i \right\rbrace$, and the relations among them are given by
%\begin{equation}
% C^a =  f^a (P) =  \left\lbrace \rho W \text{,  } (h W^2 + B^2 ) v^i - (v\cdot B) B^i \text{,  } h W^2 - p + B^2 - \frac{1}{2} ((v\cdot B)^2 + \frac{B^2}{W^2})  \right\rbrace
%\end{equation}
\begin{eqnarray*}
D &=& \rho W ~, \\
S^i &=& (h W^2 + B^2 ) v^i - (v\cdot B) B^i ~,\\
U &=& h W^2 - p + B^2 - \frac{1}{2} \left[ (v\cdot B)^2 + \frac{B^2}{W^2} \right]  ~,\\
B^i &=& B^i~,
\end{eqnarray*}
where $W= (1-v^2 )^{-1/2}$ is the Lorentz factor, the pressure $p$ is defined through an equation of state and the enthalpy is defined by, $h:= \rho (1 + \epsilon) + p$ .
The evolution system is written,
\begin{eqnarray}
 && \partial_t D + \partial_k N^k = 0 ~; \quad  N^k = D v^k ~,
\label{evol_D} \\
 && \partial_t S^i + \partial_k T^{ki} = 0 
\label{evol_S} ~, \\ %\text{  ;} \quad S^{ki} = h W^2 v^i v^k + \gamma^{ki} \left[ p+(E^2 + B^2 )/2 \right] - E^i E^k - B^i B^k \\ 
 && \partial_t U + \partial_k S^{k}   = 0  
\label{evol_U} ~,\\
 && \partial_t B^i + \partial_k M^{ki} = 0 ~; \quad M^{ki} = 2 \, B^{[i} v^{k]}  ~,
\label{evol_B} 
\end{eqnarray}
where
\begin{equation}
 T^{ki} = h W^2 v^k v^i - E^k E^i - B^k B^i  + \delta^{ki} \left[ p+ \frac{1}{2}(E^2 + B^2 ) \right] ~. \nonumber
%	&=& v^{(i} S^{k)} - (v\cdot B) \, v^{(i} B^{k)} - \frac{B^i B^k}{W^2} + \gamma^{ki} \left[ p+ \frac{1}{2}(B^2 - E^2 ) \right] 
\end{equation}
Since the electric field can be obtained from the ideal MHD condition $E^i = -\epsilon^{ijk} \, v_j \, B_k $, % = \frac{\epsilon^{ijk} B_j S_k }{hW^2 + B^2} 
%\begin{equation*}
% E^2 = \frac{S^2 B^2 - (S\cdot B)^2}{(hW^2 + B^2 )^2}
%\end{equation*}
all the previous fluxes can be easily written in terms of the primitive fields. 
Finally, we shall define $\mathcal{E}:= hW^2$ and $\Theta := \mathcal{E} + B^2$, in order to simplify our following calculations.

%\DV{Repasar con cuidado los $ki$ (lo que usamos en todo el paper) y $ik$, Ricard ha encontrado un typo que hemos corregido, pero mirar en todo el paper.}
The filtered version of the system can be written as: 
\begin{eqnarray}
	\partial_t \overline{D} + \partial_k N^k (\widetilde{P}) &=&  \partial_k \overline{\tau}^{k}_{N} \label{eq:rmhd_d} \\
	\partial_t \overline{S}^i + \partial_k T^{ki}(\widetilde{P}) &=&  \partial_k \overline{\tau}^{ki}_{T} \\
	\partial_t \overline{U} + \partial_k S^{k}(\widetilde{P}) &=&  \partial_k \overline{\tau}^{k}_{S} \\
	\partial_t \overline{B}^i + \partial_k M^{ki}(\widetilde{P}) &=& \partial_k \overline{\tau}^{ki}_{M} \label{eq:rmhd_b}
\end{eqnarray}
where on the right-hand side we have introduced the SFS tensors associated to each flux, as defined above.

As in the non-relativistic case, we propose to model the filtered SFS terms appearing in the equations \eqref{eq:rmhd_d}-\eqref{eq:rmhd_b} by means of a compact application of the gradient model. First of all, we define the {\em double gradient operator} $H$, acting on any given field $X$, as %$\tau(X):= -\xi \, \nabla \frac{dX}{d\overline{C}^b} \cdot \nabla \overline{C}^b$. 
\begin{equation}
 H_{X} = H(X):= \nabla \frac{dX}{d\overline{C}^b} \cdot \nabla \overline{C}^b~,
\end{equation}
which satisfies a sort of generalized Leibniz's rule, 
\begin{equation}\label{leibniz}
 H(X Y) =  X \, H(Y) + Y \,  H(X) + 2 \nabla X \cdot \nabla Y~.
\end{equation}
Notice that, when acting on any conserved field, it vanishes by its definition (i.e., $H_D = H_{S}^i = H_U = H_{B}^i = 0$). On the other hand, when the operator applies to a non-conserved variable, the quantity is non-zero. This holds in particular for $\{\widetilde{p},\widetilde{\Theta},\widetilde{v}^k\}$ as we shall see below.

By applying these rules and using eq.~\eqref{tau-alt}, the SGS gradient tensors approximating the SFS terms of eqs.~\eqref{eq:rmhd_d}-\eqref{eq:rmhd_b} read:
\begin{eqnarray}
  \tau^{k}_{N}  = -\xi \, H_{N}^k ~~&,&~~
  \tau^{ki}_{T} = -\xi \, H_{T}^{ki} \nonumber \\
  \tau^{k}_{S}  = 0  ~~&,&~~
  \tau^{ki}_{M} = -\xi \, H_{M}^{ki}  \label{eq:sgs_gradient}
\end{eqnarray}
where the set of the $H$ tensors, after some algebraic manipulations, can be written as:\footnote{We use a mixed notation for scalar products when it comes about the gradients, i.e. $\nabla X\cdot\nabla Y$ instead of $\nabla_i X \nabla^i Y$, in order to make well visible the gradient terms, which are the core of the SGS model and always appear contracted to each other.}
\begin{eqnarray}
 H^{k}_{N} &=&  2 \, \nabla \overline{D} \cdot \nabla \widetilde{v}^k + \overline{D} \, H^{k}_v ~, \label{tau_N}\\ 
 && \nonumber\\ 
 H^{ki}_{T} &=& 2 \left[ \nabla \widetilde{\mathcal{E}} \cdot \nabla (\widetilde{v}^i \widetilde{v}^k ) + \widetilde{\mathcal{E}} \left( \widetilde{v}^{(i} H_{v}^{k)} +  \nabla \widetilde{v}^{i} \cdot \nabla \widetilde{v}^{k} \right) \right] \nonumber \\
&+&  \widetilde{v}^i \widetilde{v}^k H_{\mathcal{E}} - 2\left[ \nabla \overline{B}^{i} \cdot \nabla \overline{B}^{k} + \nabla \widetilde{E}^{i} \cdot \nabla \widetilde{E}^{k} + \widetilde{E}^{(i} H_{E}^{k)}   \right]  \nonumber \\
&+& \delta^{ki} \left[ H_p + \nabla \overline{B}_{j} \cdot \nabla \overline{B}^{j} + \nabla \widetilde{E}_{j} \cdot \nabla \widetilde{E}^{j} + \widetilde{E}_{j} H_{E}^{j} \right]~, \label{tau_T} \\
% H^{ki}_{T} &=& 2 \left[ \nabla \widetilde{\mathcal{E}} \cdot \nabla (\widetilde{v}^i \widetilde{v}^k ) + \widetilde{\mathcal{E}} \left( \widetilde{v}^{(i} H_{v}^{k)} +  \nabla \widetilde{v}^{i} \cdot \nabla \widetilde{v}^{k} \right) +  \widetilde{v}^i \widetilde{v}^k H_p \right] \nonumber \\
%&-& 2\left[ \nabla \overline{B}^{i} \cdot \nabla \overline{B}^{k} + \nabla \widetilde{E}^{i} \cdot \nabla \widetilde{E}^{k} + \widetilde{E}^{(i} H_{E}^{k)}   \right]  \label{tau_T} \\
%&+& \left( \delta^{ki} -  \widetilde{v}^i \widetilde{v}^k \right)  \left[ H_p + \nabla \overline{B}_{j} \cdot \nabla \overline{B}^{j} + \nabla \widetilde{E}_{j} \cdot \nabla \widetilde{E}^{j} + \widetilde{E}_{k} H_{E}^{k} \right] \nonumber \\
&& \nonumber\\
H^{ki}_{M} &=&  4 \, \nabla \overline{B}^{[i} \cdot \nabla \widetilde{v}^{k]}   + 2 \overline{B}^{[i} H_{v}^{k]}  ~, \label{tau_M} \\ 
&& \nonumber
\end{eqnarray}
where $H_{E}^i $ is just the Hodge dual of $ H^{ij}_{M}$, i.e. $H_{E}^i = \frac{1}{2} \epsilon^{i}_{\phantom ijk } H_{M}^{jk}$. Notice the values of the double gradient appearing above, $\{H_p,H_\Theta,H_v^k,H_{E}^k,H_{\mathcal{E}} \}$, are meant to approximate the SFS residuals related to the non-conserved fields,  $\{\bar{\tau}_{p},\bar{\tau}_{\Theta},\bar{\tau}_v^k,\bar{\tau}_E^k,\bar{\tau}_{\mathcal{E}}\}$, defined according to equation \eqref{eq:sfs_auxiliary_field} as $\bar{\tau}_{X} \backsimeq -\xi H_X$ (exactly like for the conserved field SFS residuals).
%
%\DV{Habia puesto en algun momento una forma explicita de $E_jH_E^j$, que es sencilla, disfrutando la identidad del producto de dos Levi-Civita. La pondria de nuevo (se ha borrado me parece...), facilita la implementacion. Me parece bien tenerla en el texto por tema de elegancia solo, pero en la LES no creo... Ricard tiene implementada la $H_E$ pero no hace falta, ocupa memoria y no gana nada, considerando que aparece solo 2 veces, y la contraccion con $E_j$ simplifica las cosas.}
%
%Notice that $H_{\mathcal{E}}$ can be easily obtained from the energy relation, i.e. $U = \mathcal{E} - p + \frac{1}{2}(E^2 + B^2 )$, yielding $H_{\mathcal{E}}= H_p - \nabla \overline{B}_{j} \cdot \nabla \overline{B}^{j} - \nabla \widetilde{E}_{j} \cdot \nabla \widetilde{E}^{j} - \widetilde{E}_{k} H_{E}^{k}$.
%
%\CPIn the following calculations we have also used that it is always possible to take three orthogonal vectors, like e.g. $\left\lbrace v^i , b^i , E^i \right\rbrace $ (with $ b^i := B^i - \frac{(v\cdot B)}{v^2} v^i$ ) and decompose the spatial metric in this basis.
%	Expanding, one gets the following relation:}
%\CP{Fede, can you discuss how this is being used in the last equations}
%\begin{equation*}
%E^i E^k = \gamma^{ki} E^2 - B^2 v^i v^k -v^2 B^i B^k  + 2 (v\cdot B) v^{(i} B^{k)}
%\end{equation*}
%
Their explicit expressions are obtained by computing the following set of equations in the order in which they appear, where the quantities $\widetilde{\Psi}$ denote the auxiliary fields which are used to simplify the presentation (and also to facilitate their implementation):
\begin{widetext}
\begin{eqnarray}
\widetilde{\Psi}_{v}^k &=& \frac{2}{\widetilde{\Theta}} \left\lbrace \nabla (\widetilde{v}\cdot \overline{B}) \cdot \nabla \overline{B}^k  - \nabla \widetilde{\Theta} \cdot \nabla \widetilde{v}^k   
+ \frac{\overline{B}^k}{\widetilde{\mathcal{E}}} \left[  \widetilde{\Theta} \nabla \overline{B}_j \cdot \nabla \widetilde{v}^j + \overline{B}_j \nabla \overline{B}^j \cdot \nabla (\widetilde{v}\cdot \overline{B}) - \overline{B}_j \nabla \widetilde{v}^j \cdot \nabla \widetilde{\Theta} \right]  \label{hTauv}  \right\rbrace ~, \nonumber  \\
\widetilde{\Psi}^{ki}_{M} &=& \frac{4}{\widetilde{\Theta}} \left[  \widetilde{\Theta} \, \nabla \overline{B}^{[i}  \cdot \nabla \widetilde{v}^{k]} +  \overline{B}^{[i} \nabla \overline{B}^{k]} \cdot \nabla (\widetilde{v}\cdot \overline{B}) - \overline{B}^{[i} \nabla \widetilde{v}^{k]} \cdot \nabla \widetilde{\Theta} \right] ~,
 \nonumber \\
\widetilde{\Psi}_{\Theta} &=&  \frac{\widetilde{\Theta}}{\widetilde{\Theta} -\widetilde{E}^2} \left\lbrace \nabla \overline{B}_{j} \cdot \nabla \overline{B}^{j} - \nabla \widetilde{E}_{j} \cdot \nabla \widetilde{E}^{j} - \overline{B}_{[i}\widetilde{v}_{k]} \, \widetilde{\Psi}^{ki}_{M} \right\rbrace ~, \nonumber \\
\widetilde{\Psi}_{A}  &=& \widetilde{W}^2 \left( \widetilde{p} \, \frac{d\widetilde{p}}{d\widetilde{\epsilon}} + \widetilde{\rho}^2 \, \frac{d\widetilde{p}}{d\widetilde{\rho}} \right) ~, \nonumber 
\end{eqnarray}
\begin{eqnarray}
\frac{H_{\rm p}}{\widetilde{\Theta}- \widetilde{E}^2 } &=&  \frac{\widetilde{\mathcal{E}} \, \widetilde{W}^2}{(\widetilde{\rho} \, \widetilde{\mathcal{E}} - \widetilde{\Psi}_{A})(\widetilde{\Theta} - \widetilde{E}^2 ) \widetilde{W}^2 + \widetilde{\Psi}_{A} \, \widetilde{\Theta}} \left\lbrace \widetilde{\rho} \left( \nabla \frac{d\widetilde{p}}{d\widetilde{\rho}} \cdot \nabla \widetilde{\rho} + \nabla \frac{d\widetilde{p}}{d\widetilde{\epsilon}} \cdot \nabla \widetilde{\epsilon} \right)  - 2 \frac{d\widetilde{p}}{d\widetilde{\epsilon}} \, \nabla \widetilde{\rho} \cdot \nabla \widetilde{\epsilon}  \right. \nonumber\\
&-&  \left.  \left(\widetilde{\mathcal{E}} \frac{d\widetilde{p}}{d\widetilde{\epsilon}} - \widetilde{\Psi}_{A}\right) \left[ \frac{\widetilde{W}^2}{4} \nabla \widetilde{W}^{-2} \cdot \nabla \widetilde{W}^{-2} + \nabla \widetilde{W}^{-2} \cdot \nabla (\ln \widetilde{\rho}) \right]  -  \frac{2}{\widetilde{W}^2}\frac{d\widetilde{p}}{d\widetilde{\epsilon}} \left[   \nabla \overline{B}_j \cdot \nabla \overline{B}^j +  \nabla \widetilde{W}^{2} \cdot \nabla \widetilde{h} \right]  \right. \label{tau_p} \\
&-&  \left.  \left(\widetilde{\mathcal{E}} \frac{d\widetilde{p}}{d\widetilde{\epsilon}} + \widetilde{\Psi}_{A}\right) \left[ \widetilde{v}_k \widetilde{\Psi}_{v}^k +  \nabla \widetilde{v}_{j} \cdot \nabla \widetilde{v}^{j} + \widetilde{W}^2 \, \nabla \widetilde{W}^{-2} \cdot \nabla \widetilde{W}^{-2} \right]  +   \frac{1}{\widetilde{\mathcal{E}}} \left[ \left(\widetilde{\mathcal{E}} \frac{d\widetilde{p}}{d\widetilde{\epsilon}} + \widetilde{\Psi}_{A}\right)(\widetilde{\Theta}- \widetilde{E}^2 ) - \frac{\widetilde{\Psi}_{A} \, \widetilde{\Theta}}{\widetilde{W}^2}   \right] \frac{\widetilde{\Psi}_{\Theta}}{\widetilde{\Theta}}  \right\rbrace ~, \nonumber \\
H_{\mathcal{E}} &=& H_p - \nabla \overline{B}_{j} \cdot \nabla \overline{B}^{j} - \nabla \widetilde{E}_{j} \cdot \nabla \widetilde{E}^{j} - \widetilde{E}_{k} H_{E}^{k}  ~,\\
H_{\Theta} &=& \widetilde{\Psi}_{\Theta} + \frac{\widetilde{\Theta}}{\widetilde{\Theta} -\widetilde{E}^2} H_p ~, \label{tau_Theta} \\ 
H_{v}^k &:=& \widetilde{\Psi}_{v}^k - \left( \widetilde{v}^k + \frac{\widetilde{v}\cdot \overline{B}}{\widetilde{\mathcal{E}}} \overline{B}^k \right)  \frac{H_{\Theta}}{\widetilde{\Theta}} ~. \label{Tauv}
\end{eqnarray}
\end{widetext}
Note that these equations are remarkably much more involved than in the well-known non-relativistic case \cite{paper1}. %The only terms which are basically the same in both cases are $\tau_{\rm ind}$ and $\tau_M$. \FC{este statement no me queda claro, por que decimos eso?} \DV{Aqui, y abajio, estoy relacionando cosas bien conocidas, el caso non relativista, donde hay muchos papers de literatura, con nuestro caso, para dar alguna referencia y correspondencia. } 
%The SFS terms arising from the momentum equation, finally, are extremely more complicated.
The main difference being, of course, the more complicate relationship between the conserved and primitive variables in the relativistic setting. 
Notice that the continuity equation now acquires a SFS term, while the energy equation does not contain any SFS residual. Such ``exchanged roles'' of the continuity and energy equations comes from the non-relativistic limit and can be seen already at the level of the equations. %\FC{ Esta ``inversion de roles'' entre la eqn de continuidad y la de energia venia en realidad del proceso de tomar el limite no-relativista. No me queda claro si quedo bien expresado asi..} \DV{De nuevo, estoy comparando con algo conocido... Pero bueno, si quieres refrasearlo, adelante.}
%\CP{añadi una referencia al paper I y ya esta, asi se ve la correspondencia}
	
Notice also that, depending on the equation of state, some terms can be considerably simplified. For instance, $\widetilde{\Psi}_{A} = W^2\Gamma p\rho$ for an ideal gas EoS with coefficient $\Gamma$.

%\DV{No se puede obtener algun limite interesante non-relativistico? }

\section{Application to magnetic relativistic simulations}\label{sec:simulations}

One way to check the validity of the relativistic SGS gradient model is by performing a detailed analysis of numerical simulations displaying turbulence. Therefore, we will consider a decaying (i.e., non-forced) turbulent dynamics in relativistic MHD, triggered by the KHI, a scenario already studied especially in the non-relativistic case \cite{obergaulinger10,beckwith11}. Notice that the KHI provides different stages of the turbulent flow: the initial development at small scales, the transfer of kinetic to magnetic energy, the saturation and mixing, and the final slow decay \cite{obergaulinger10,paper1}. Moreover, the KHI is thought to take place in binary neutron star mergers, and will be the natural mechanism to be studied in the forthcoming general relativistic simulations.

\subsection{Setup}\label{sec:setup}

The initial setup is just an extension from the non-relativistic \cite{beckwith11,paper1} to the special relativistic case. We evolve eqs.~\eqref{evol_D}-\eqref{evol_B}
by using the SAMRAI infrastructure \cite{hornung02,gunney16} with a code generated by the platform {\it Simflowny} \cite{arbona13,arbona18}. The numerical schemes are the same used in Ref.~\cite{paper1}, which were described in detailed in \cite{palenzuela18,vigano19}.
The discretization of the continuum equations is performed by using the Method of Lines, which allows to address separately the time and the space discretization. We employ High-Resolution Shock-Capturing (HRSC) methods \cite{toro97} to deal with the possible appearance of shocks and to take advantage of the existence of weak solutions in the equations. The fluxes at the cell interfaces are calculated by combining the Lax-Friedrichs splitting \cite{shu98} with the WENO5Z \cite{BORGES20083191} high-order non-oscillatory reconstruction scheme. The time integration of the resulting semi-discrete equations is performed by using a fourth-order Runge-Kutta scheme, which ensures the stability and convergence of the solution for a small enough time step $\Delta t \leq 0.4 ~\Delta$.  

We set our problem in Cartesian coordinates, considering a periodic box $[-L/2,L/2]^3$. We shall consider different resolutions between $128^3$ and $1024^3$, and evolve the system up to $t=20$. The primitive fields read initially:

\begin{eqnarray}
&& \rho=\rho_0 + \rho_1~\sign(y)\tanh\left(\frac{|y| - y_l}{a_l}\right)~, \\
&& v_x = v_{x0} ~\sign(y)\tanh\left(\frac{|y| - y_l}{a_l}\right) + \delta v_x ~,\\
&& v_y = \delta v_y~\sign(y)  \exp\left[-\frac{(|y| - y_l)^2}{\sigma_y^2}\right] ~,\\
&& v_z = v_{z0}~\sign(y)\exp\left[-\frac{(|y| - y_l)^2}{\sigma_z^2}\right] + \delta v_z ~,\\
&& B_x = B_{x0} ~,~
 B_x = B_{y0} ~,~
 B_z = B_{z0} ~,~
 p=p_0 ~,
\end{eqnarray}
where $a_l$ is the mixing layer scale, $y_l$ is the distance of the shear layers to the plane $y=0$, $\sigma_y$ and $\sigma_z$ are the extension scale of the initial perturbation in the $y$-direction and the profile of $v_z$, respectively. The main flow is initially given by $v_{x0}$. The standard values that we consider are $L=1$, $y_l=1/4$, $\rho_0=1.5$, $\rho_1=0.5$, $a_l=0.01$, $v_{x0}=0.5$, $B_{x0}=0.001$, $B_{y0}=B_{z0}=0$, $p_0=1$ and $\sigma_z^2=0.1$. We consider an ideal gas equation of state, $p=(\gamma-1)\rho\epsilon$, with $\gamma=4/3$. The purpose of this paper is not to explore the dynamics for different parameters (see for instance \cite{obergaulinger10} for a discussion about the role of the initial values of Mach number and magnetic field).

The initial perturbation, $\delta v_i$, is a superposition of single-mode perturbation with a number of nodes $n_i \in [1,N/2]$, periodic in the boundary box: 
\begin{eqnarray}
%\delta v_x=\delta v_0 \sin(2\pi z n_x/L) \\
%\delta v_y=\delta v_0 \sin(2\pi x n_y/L) \\
%\delta v_z=\delta v_0 \sin(2\pi y n_z/L) \\
\delta v_i=\delta v_0 \sin(2\pi x_i n_i/L) ~.
\end{eqnarray}
We underline that the specific form of the initial perturbation has no influence on the asymptotic turbulent behavior, as long as we excite the entire spectrum of modes, which can be achieved easily if the modes are not the same (or multiple) to each other, $n_i \gg 1$ and $\delta v_i \ll v_{0x}$. Hereafter, we use $n_x=11$, $n_y=7$, $n_z=5$, $\delta v_x = \delta v_z = 0.01$, $\delta v_y = 0.1$, $\sigma_y^2=0.01$. 
%We stress that we use a multi-modal perturbation, with non-trivial wavenumbers (prime between them), instead of what commonly used (a given perturbation mode). This allows us to excite virtually all the spectrum of modes,
%We stress that with this non-trivial perturbation allows to excite virtually all the spectrum of modes. 
Since the KHI is known to grow faster for smaller scales, and due to the absence of physical viscosity in this test, we do not expect a numerical convergence, at least in the growth phase, since the more we refine the grid, the more fast-growing excited mode will be included. This actually reproduces the scenario of the binary neutron star mergers, where the finest available resolutions are likely still very far from being able to capture all the relevant modes, and the simulations do not show numerical convergence, in terms of total magnetic energy and its spectra. See also our previous work \cite{paper1} for more details and for the general, similar behavior in the non-relativistic case.

\begin{figure*}[t] 
	\centering
	\includegraphics[width=0.32\linewidth]{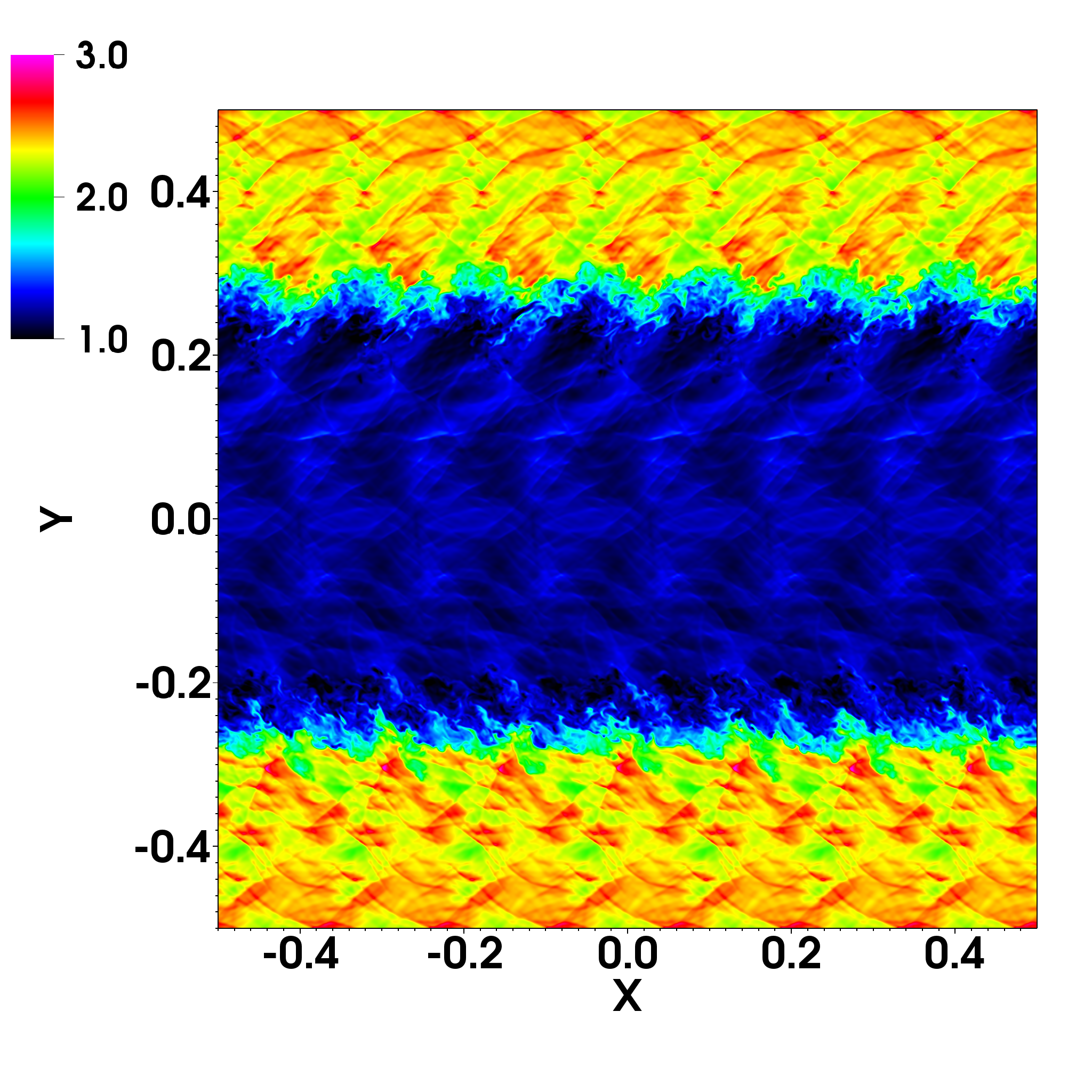}
	\includegraphics[width=0.32\linewidth]{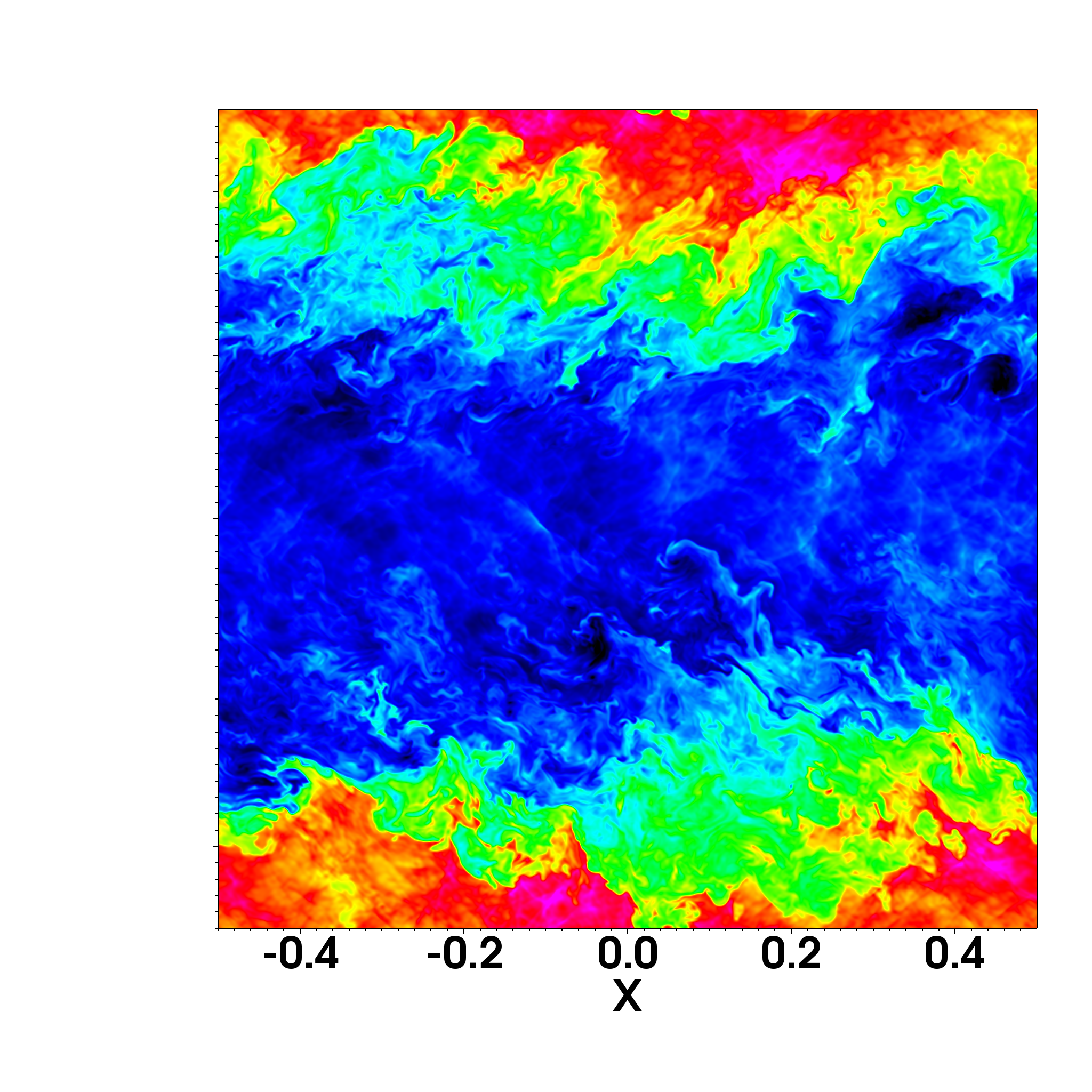}
	\includegraphics[width=0.32\linewidth]{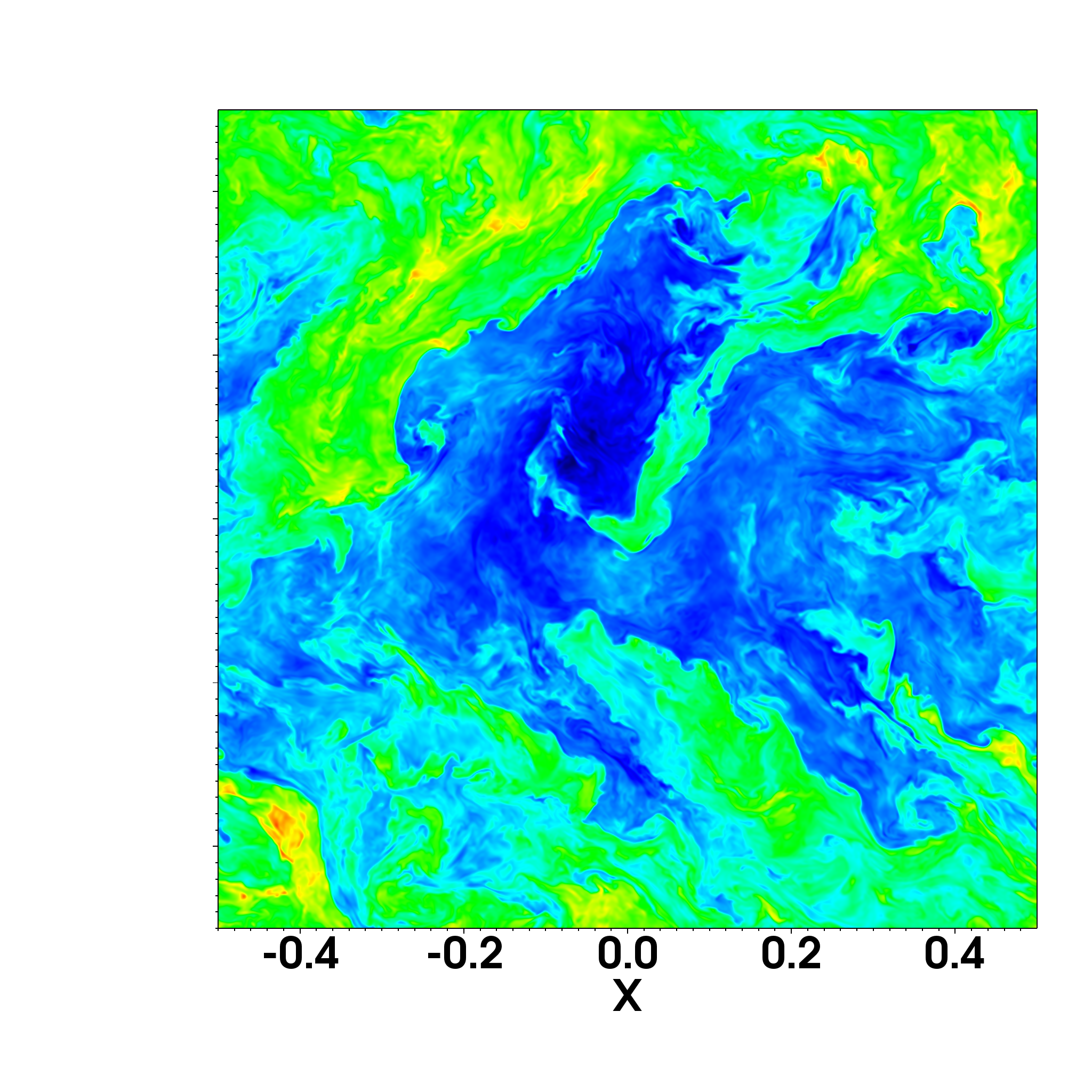}
	\caption{Evolution of the rest-mass density $\rho$ for the $N=1024^3$ case, in the $z=0$ plane, at $t=2,10,20$ (from left to right). The initial perturbations seed the initial eddies typical of the KHI, which quickly develops a turbulent non-forced dynamics.}
	\label{fig:kh3d_images}
\end{figure*}

\begin{figure}[t] 
	\includegraphics[width=0.8\linewidth]{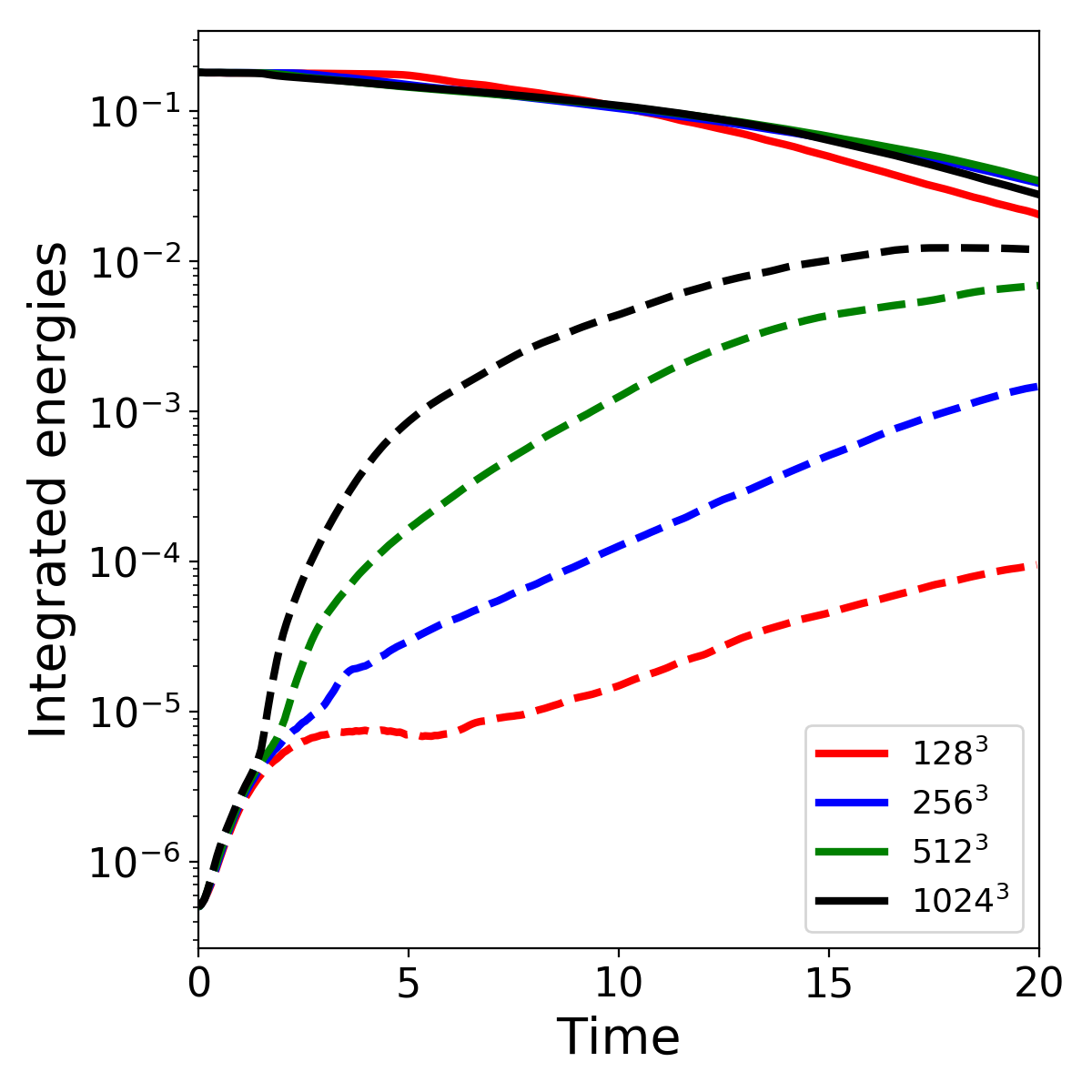}
	\caption{Evolution of the integrated energies, kinetic in solid lines and magnetic with dashed lines, for different resolutions.
		As the resolution increases, the turbulent regime develops faster and the transference from kinetic to magnetic energy is more predominant. }
	\label{fig:kh3d_evo_overall}
	% kh3d_integrated_plots_simflowny_batches.py
\end{figure}

\begin{figure*}[t] 
	\centering
	\includegraphics[width=0.32\linewidth]{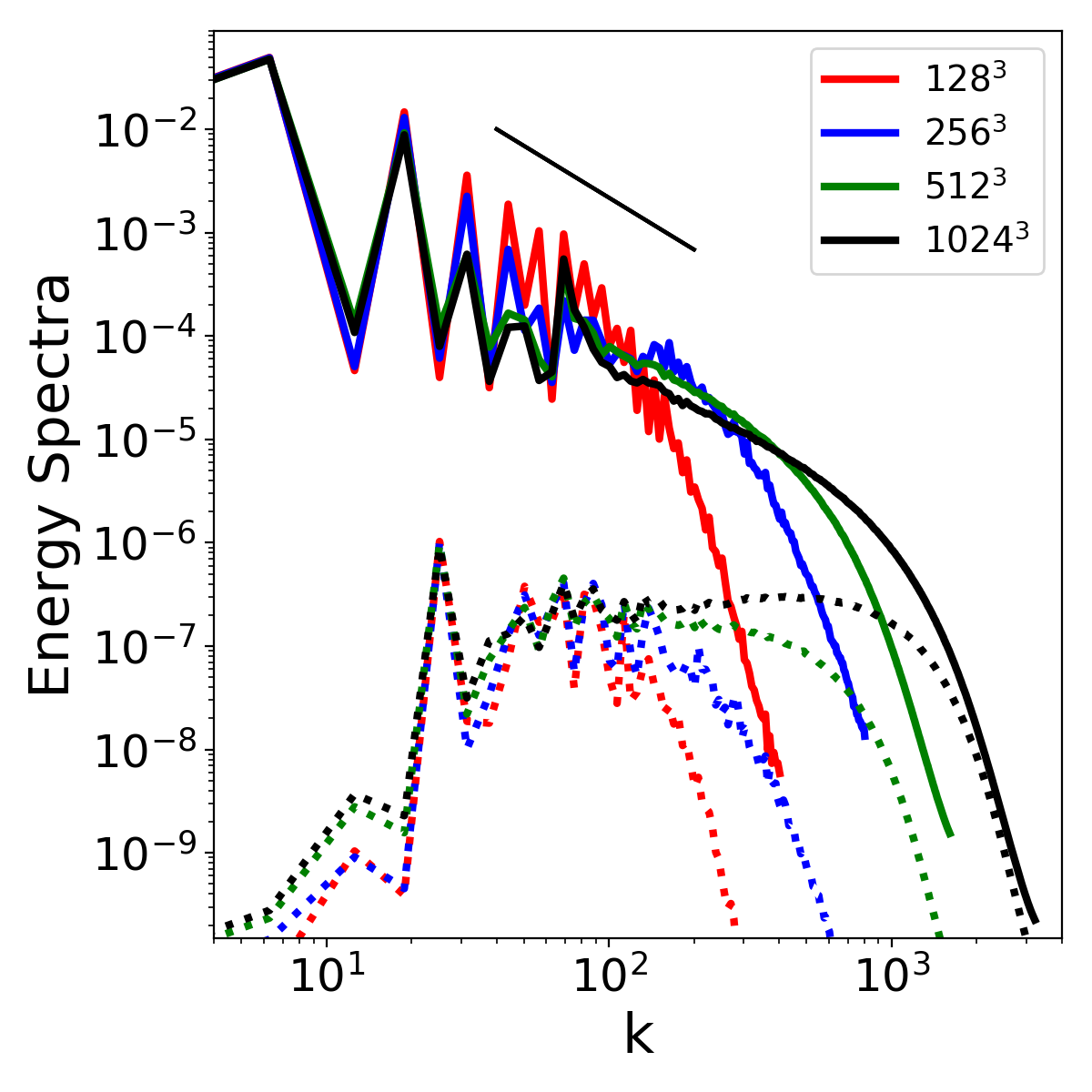}
	\includegraphics[width=0.32\linewidth]{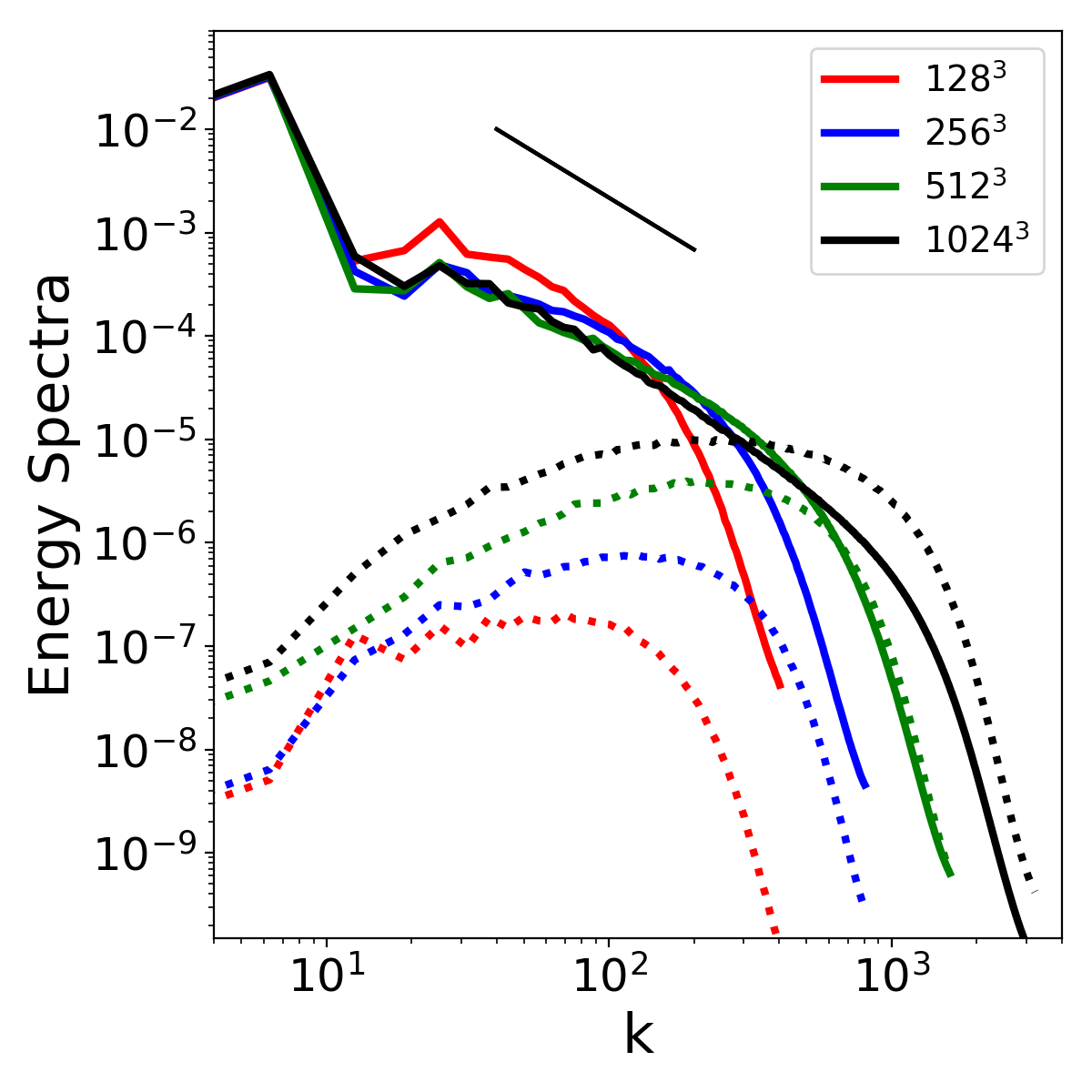}
	\includegraphics[width=0.32\linewidth]{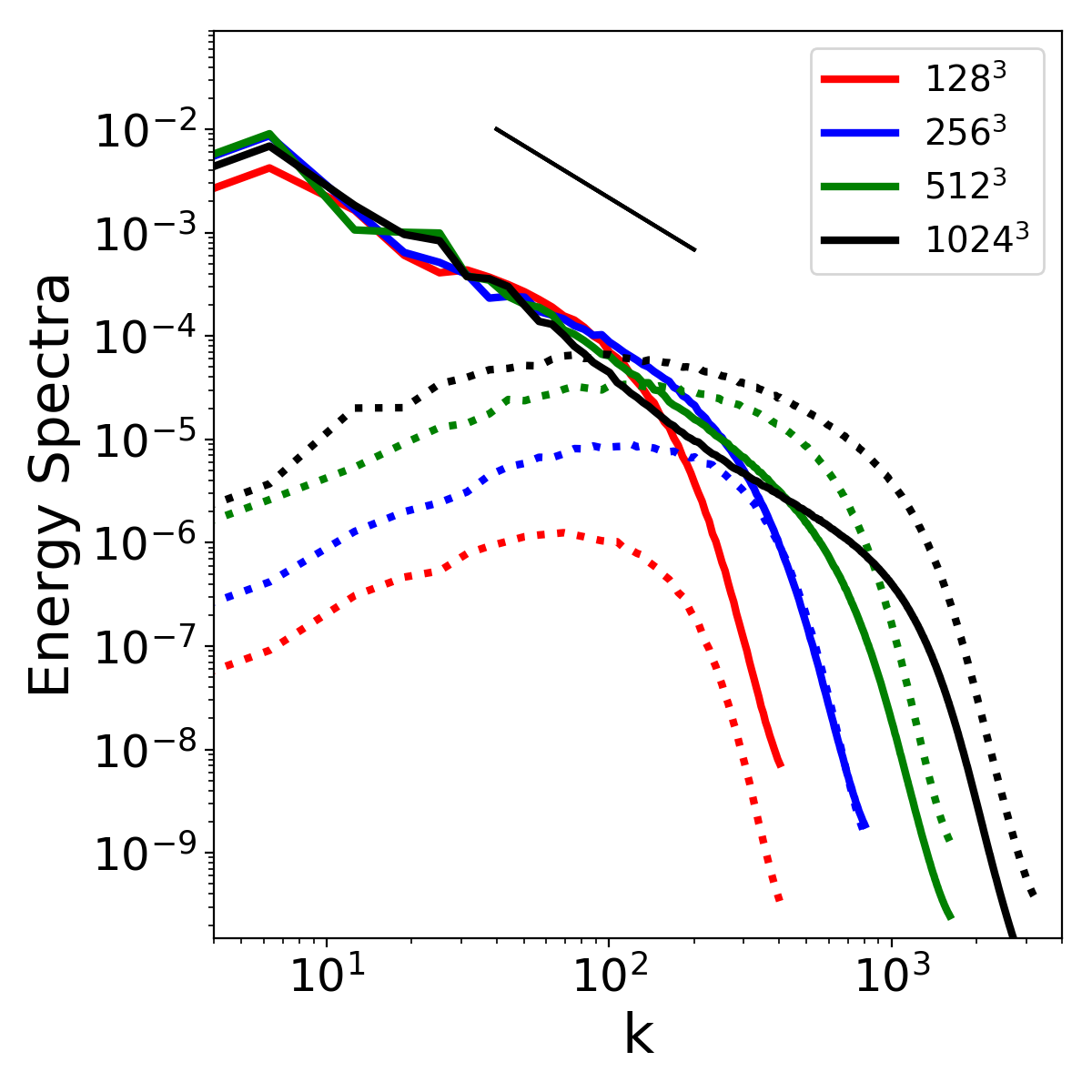}
	\caption{Spectra of the kinetic (solid) and magnetic (dashed) energies, for different resolutions, at three representative times; very early, when the turbulence is just quickly developing, an intermediate state when it is growing but at a slower pace and a late state when it saturated already (for the high resolution cases). As the resolution increases, the turbulent regime develops faster and the transference from kinetic to magnetic energy is more predominant.}
	\label{fig:kh3d_spectra}
	% python kh3d_spectral_rel_plots.py
\end{figure*}

\begin{figure}[ht] 
	\centering
	\includegraphics[width=0.49\linewidth]{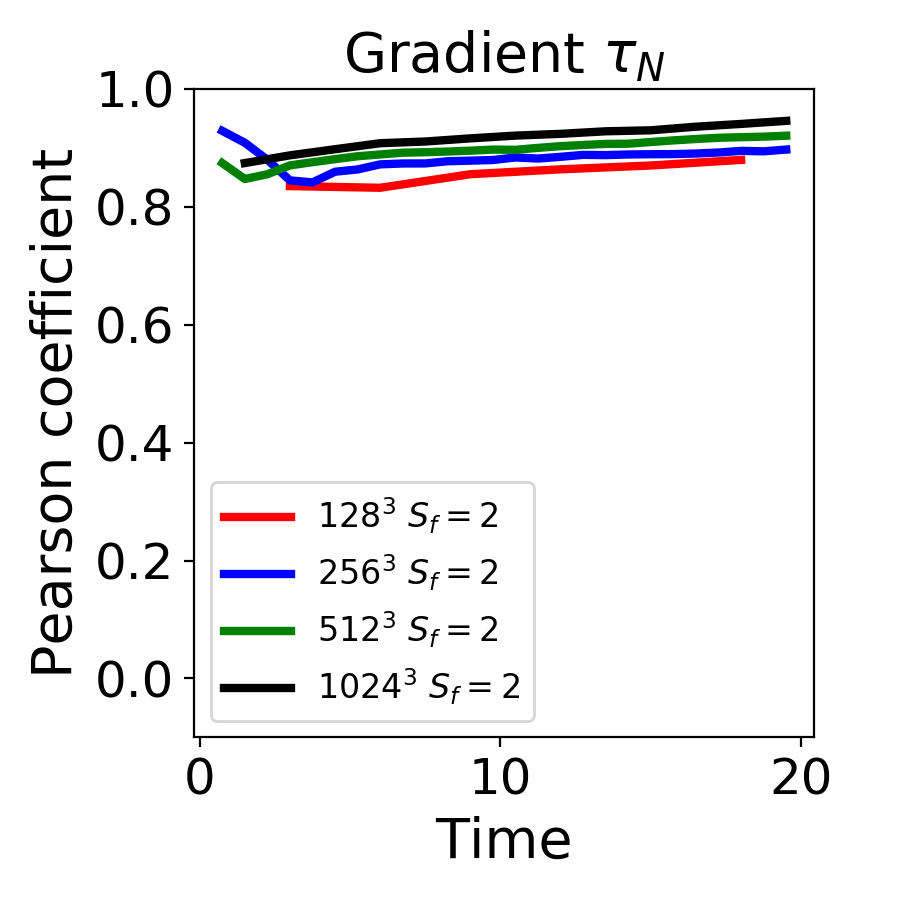}
	\includegraphics[width=0.49\linewidth]{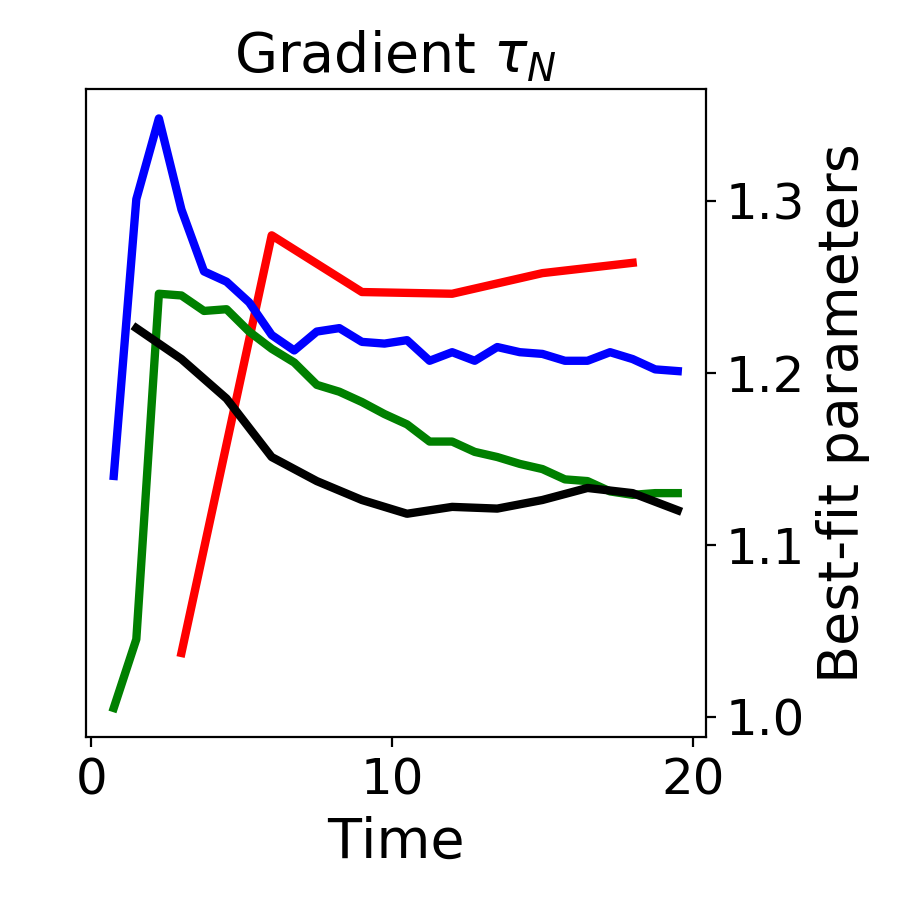}\\
	\includegraphics[width=0.49\linewidth]{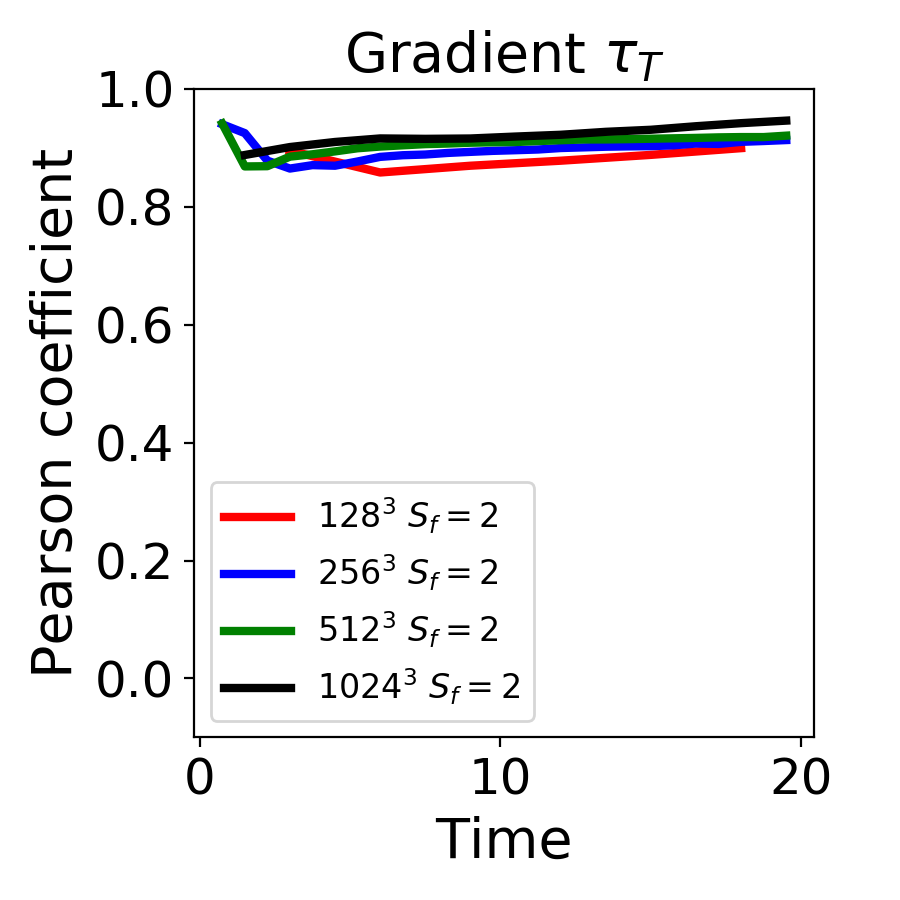}
	\includegraphics[width=0.49\linewidth]{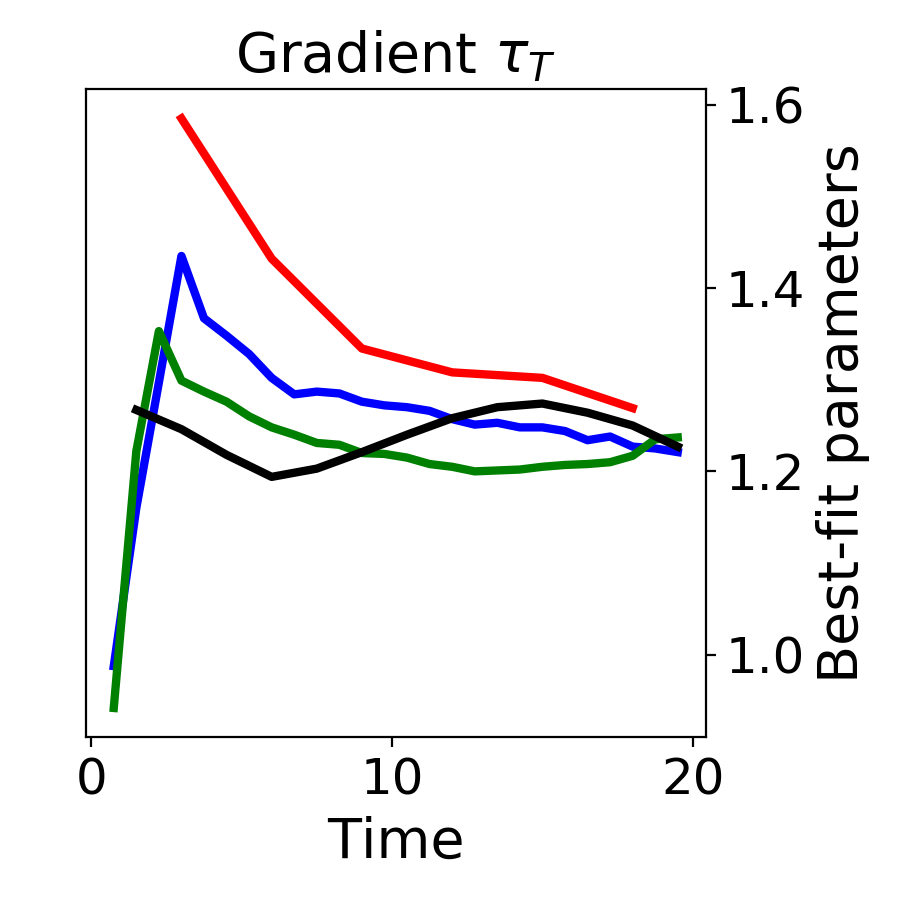}\\	
	\includegraphics[width=0.49\linewidth]{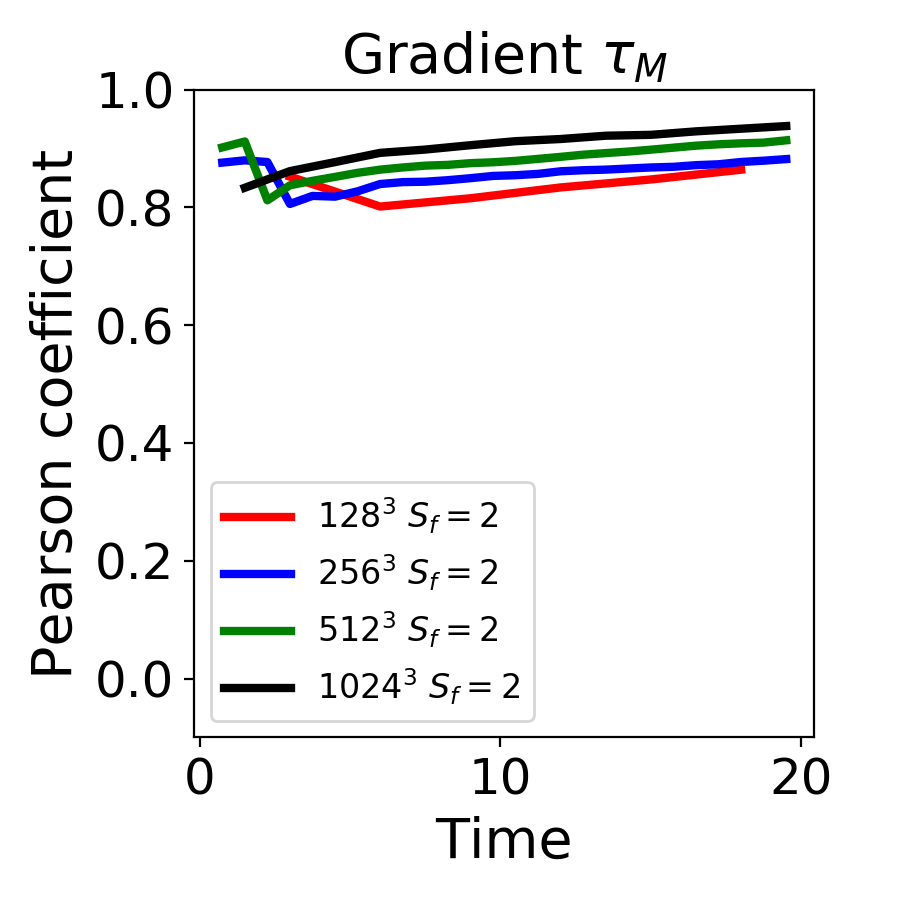}
	\includegraphics[width=0.49\linewidth]{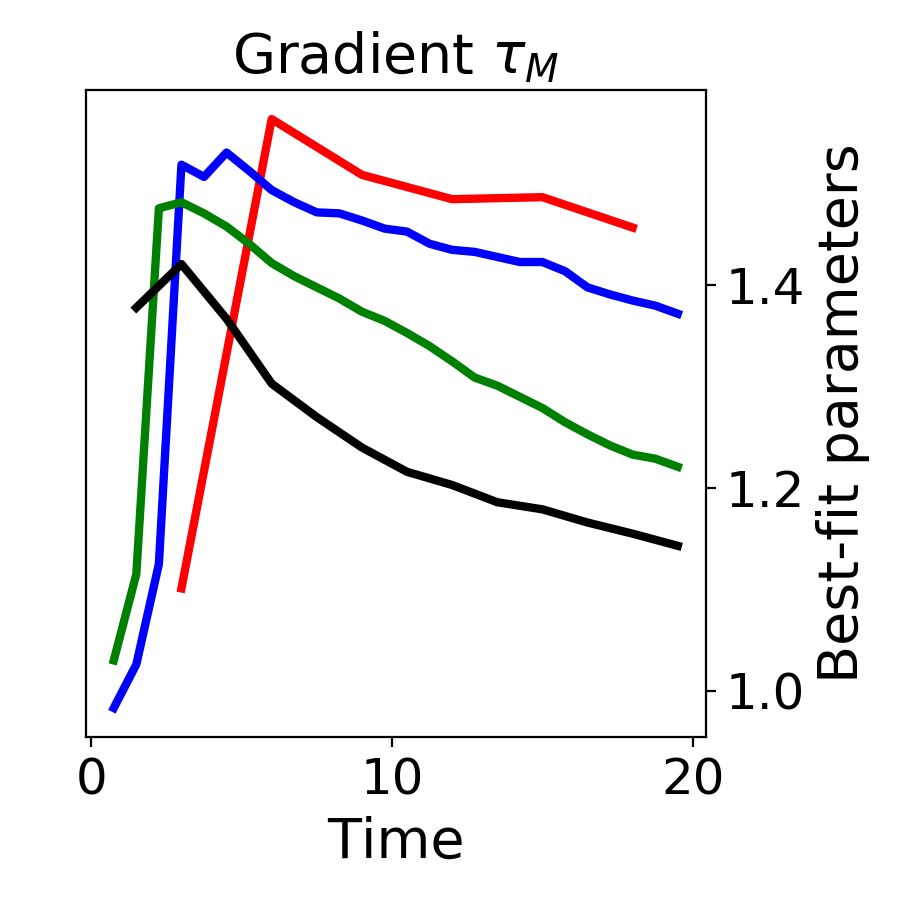}
	\caption{Evolution of the Pearson coefficients (left) and the best-fit coefficients (right) for the gradient SGS tensors $\tau_{\rm N}$ (top), $\tau_{\rm T}$ (middle) and $\tau_{\rm M}$ (bottom), for all resolutions and with a filter factor $S_f=2$.}
	\label{fig:kh3d_bestfit}
\end{figure}

\begin{figure}[ht] 
	\centering
	\includegraphics[width=0.49\linewidth]{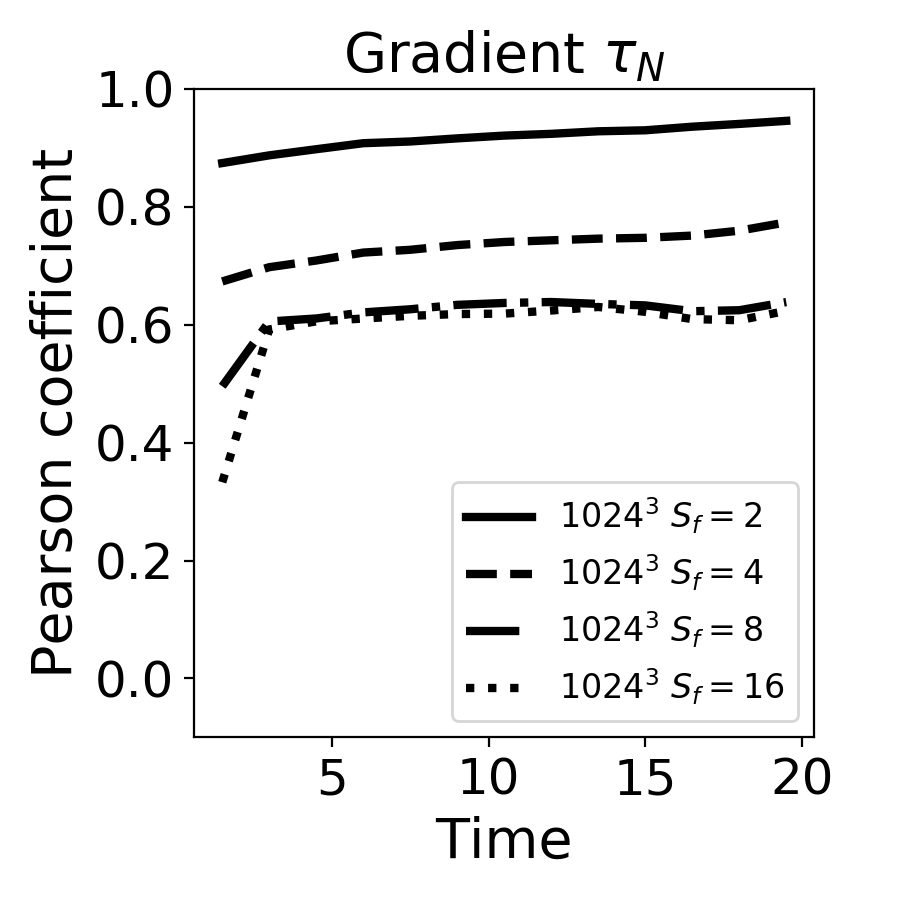}
	\includegraphics[width=0.49\linewidth]{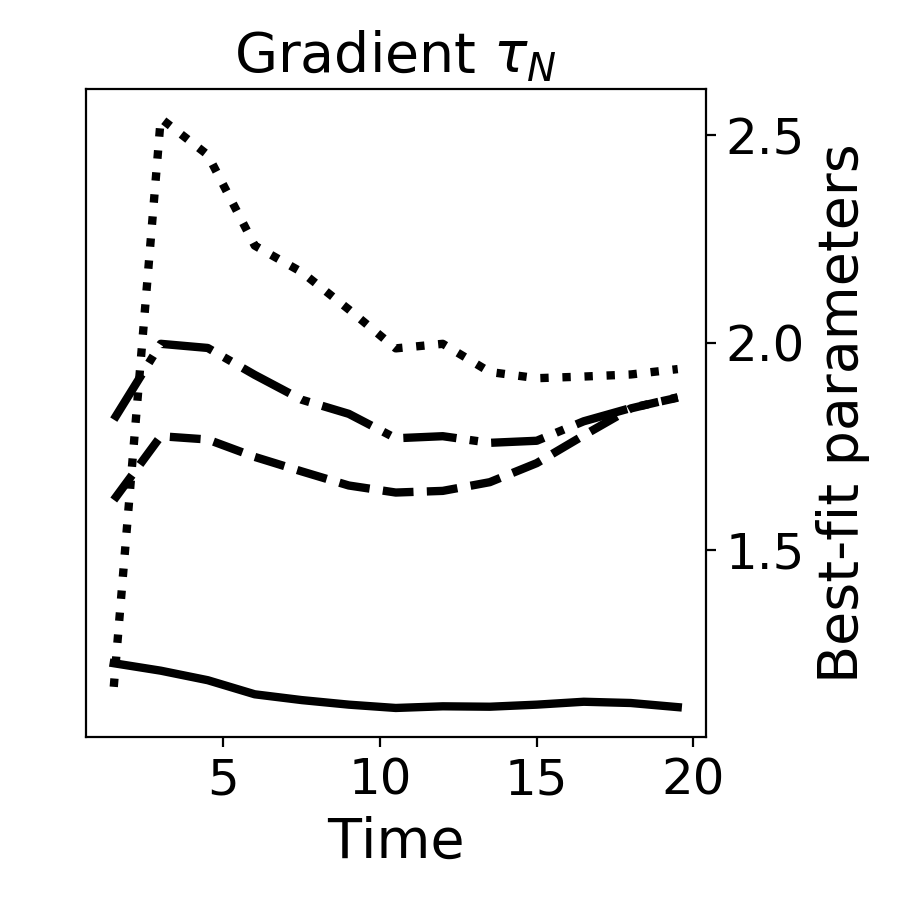}\\
	\includegraphics[width=0.49\linewidth]{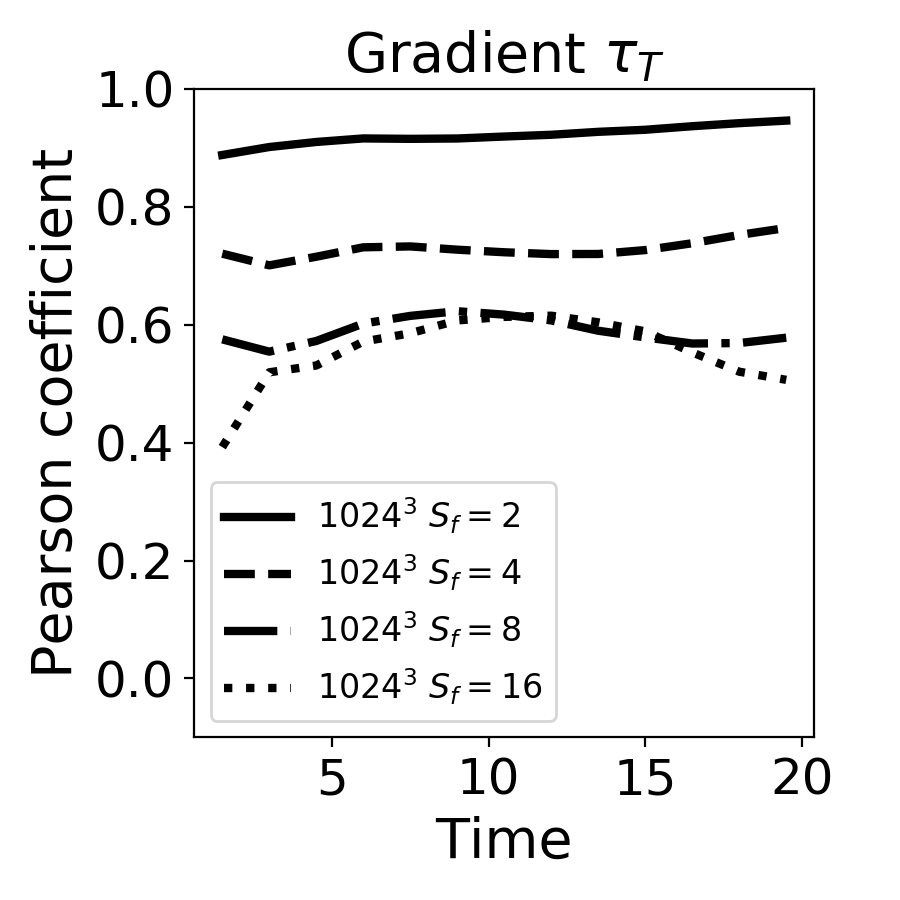}
	\includegraphics[width=0.49\linewidth]{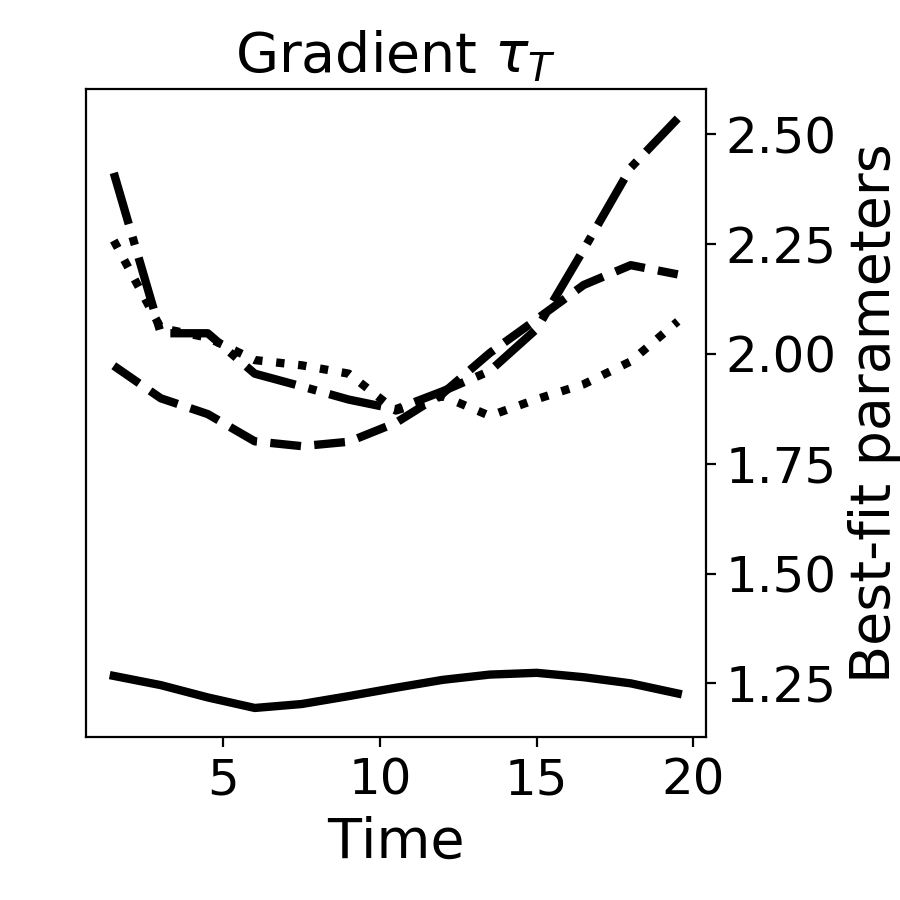}\\	
	\includegraphics[width=0.49\linewidth]{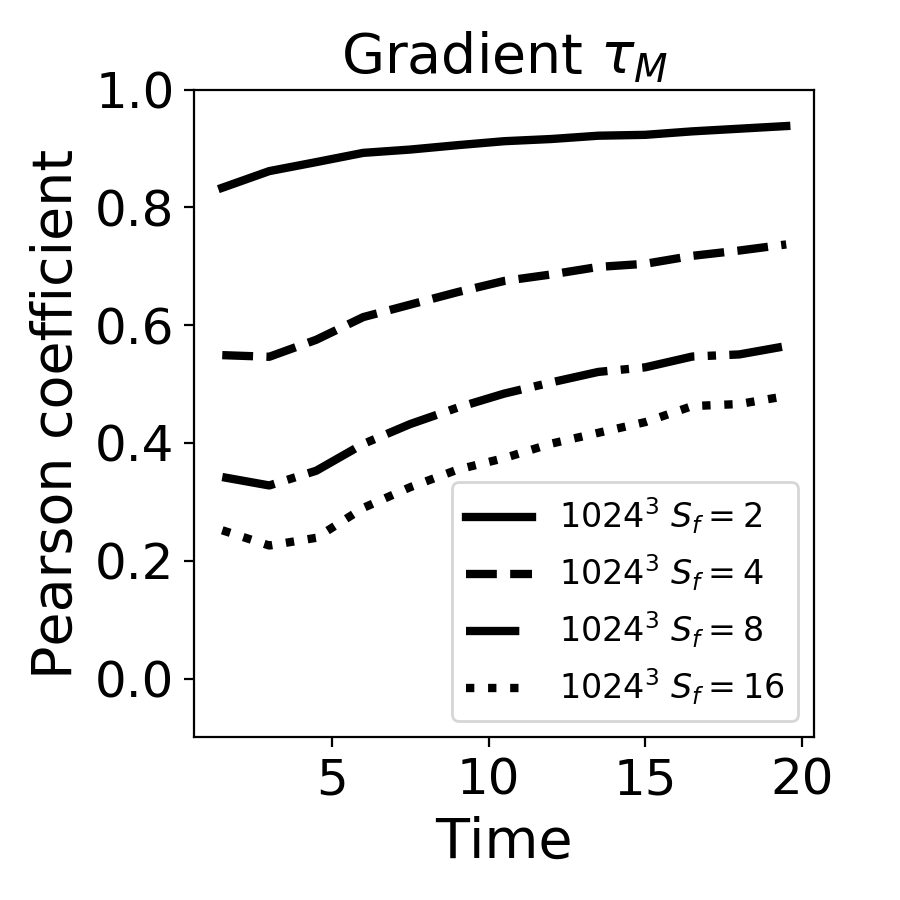}
	\includegraphics[width=0.49\linewidth]{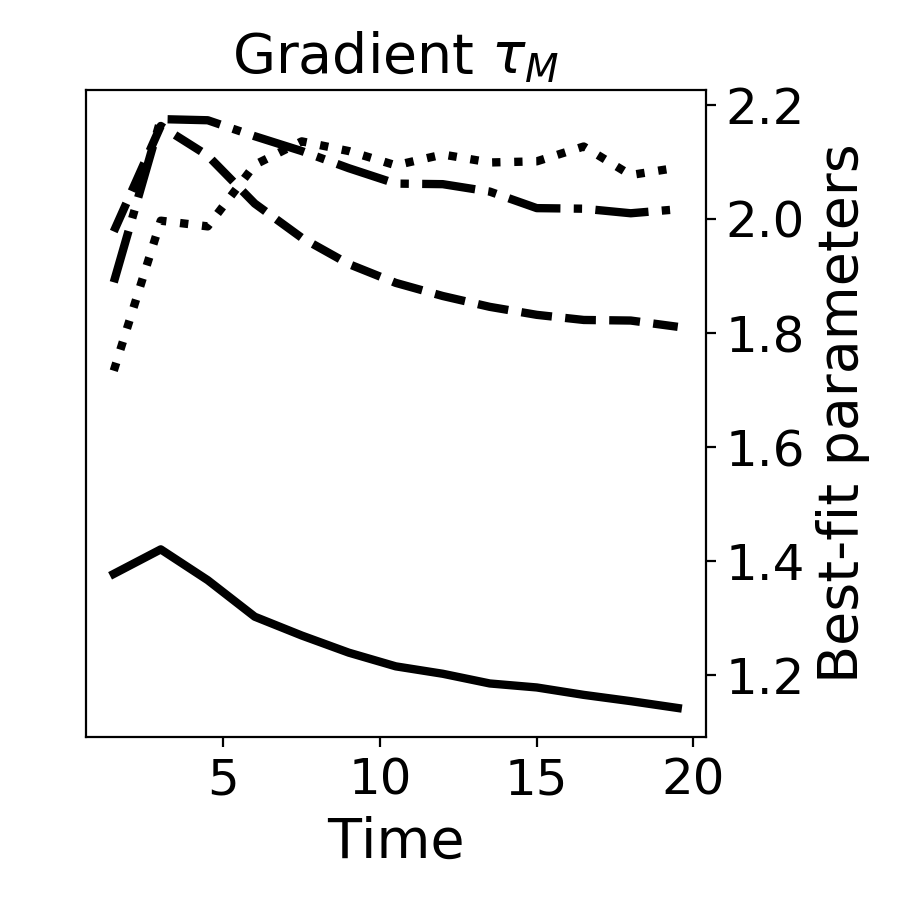}
	\caption{Evolution of the Pearson coefficients (left) and the best-fit coefficients  (right) for the gradient SGS tensors $\tau_{\rm N}$ (top), $\tau_{\rm T}$ (middle) and $\tau_{\rm M}$ (bottom), for the $N=1024^3$ case with filter factors $S_f=2,4,8,16$.}
	\label{fig:kh3d_filtersize}	
\end{figure}

\begin{figure}[ht] 
	\centering
	\includegraphics[width=0.49\linewidth]{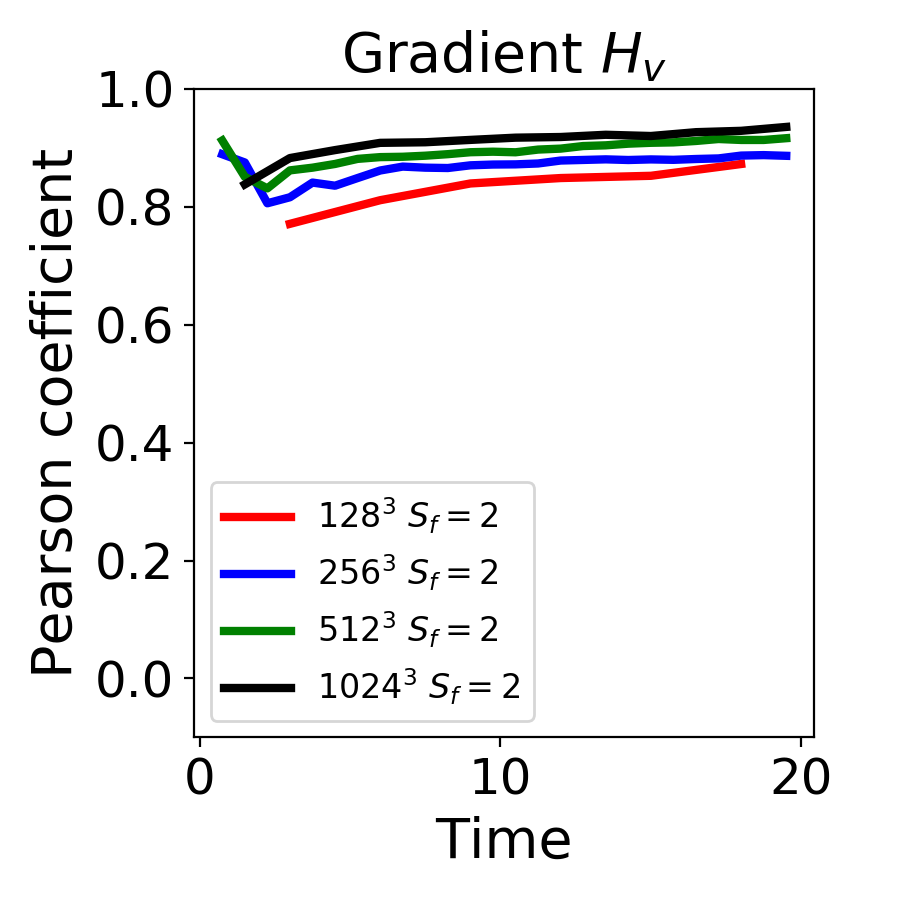}
	\includegraphics[width=0.49\linewidth]{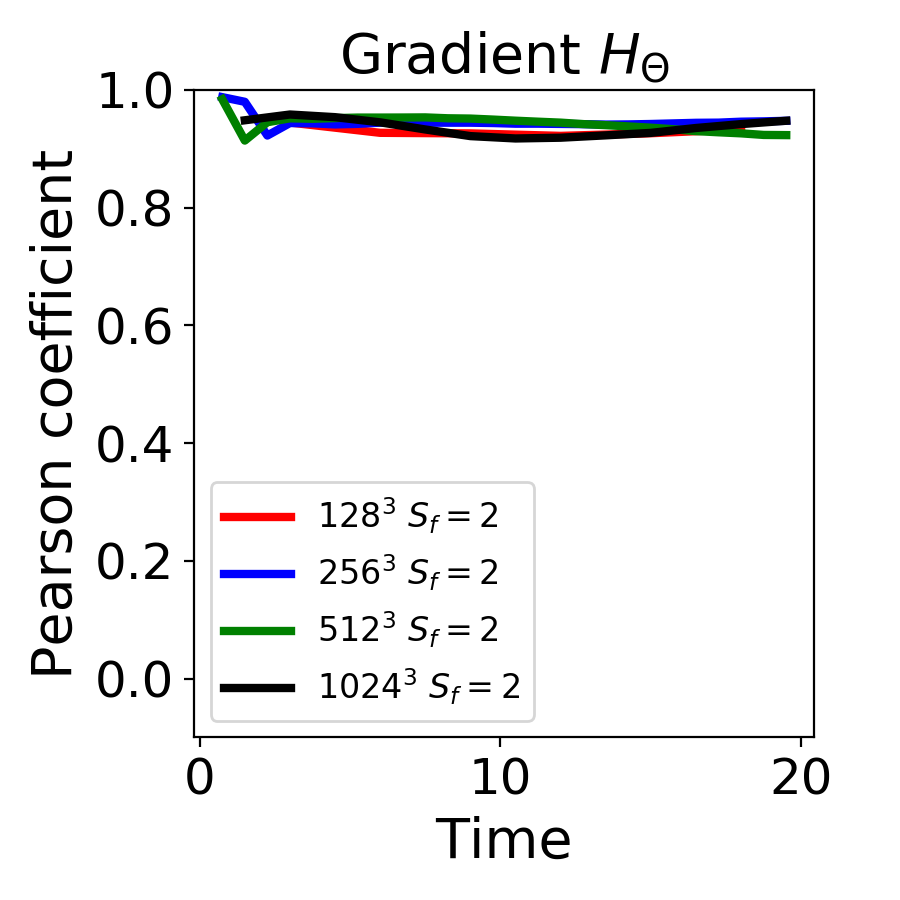}
	\caption{Evolution of the Pearson value for the correlation between the SGS term $(-\xi H_v^k)$ and the SFS term $(\widetilde{v}^k - \bar{v}^k)$ (left; as usual, we average the Pearson value over the three components $k$), and the SGS term $(-\xi H_\Theta)$ and the SFS term $(\widetilde{\Theta} - \overline{\Theta})$ (right), for all resolutions, with $S_f=2$.}
	\label{fig:kh3d_bestfit_aux}
\end{figure}

%%%%%%%%%%%%%%%%%%%%%%%%%%%%%%%%%%%%%%%%%%%%%%%%
\subsection{General behaviour}\label{sec:behaviour}
%%%%%%%%%%%%%%%%%%%%%%%%%%%%%%%%%%%%%%%%%%%%%%%%

The following qualitative behavior is similar for all the resolutions considered. The development of the turbulent dynamics is illustrated
in Fig.~\ref{fig:kh3d_images}, where the density distribution over the slice $z=0$ is displayed at times $t=\{2,10,20\}$ for the highest resolution case with $N=1024^3$. At the beginning, the instability develops as small-scale structures with modes given by the initial perturbation. The development starts at the shear layer, where the transfer of kinetic to magnetic energy (dynamo effect) is fast up to $t \sim 2$ for all the resolutions (left panel), extending up to $t \sim 4$ for the highest one. Then, a larger scale mixing takes place (middle and right panel), during which the effectiveness of the dynamo mechanism reduces. After that period the mixing is completed and the fluid looks isotropic and homogeneous.

In Fig.~\ref{fig:kh3d_evo_overall} we compare the evolution of the integrated kinetic and magnetic energy for different resolutions.
In our initially kinetic-dominated setup, both the internal and  magnetic energies grows during the evolution, at the expense of the kinetic energy. For this particular setup, the internal energy is quantitatively the dominant one (due to the chosen values of the initial pressure), while the magnetic energy is always quite smaller than the kinetic energy. 

As it has been observed already by several works, the transfer from kinetic to magnetic energy is more prominent at small scales. Therefore, since the initial perturbation spans all the scales and we have no viscosity included, when we consider higher resolution we capture more unstable modes, which in turn enhance further the magnetic growth. For the explored resolutions, we do not observe yet a saturation of the magnetic field in the homogeneous phase. Actually, in all cases, the dynamo mechanism slowly goes on, allowing in time to reduce the difference between the total kinetic and magnetic energy (possibly showing a slow approach to equipartition at the end of our $N=1024^3$ simulation). Similar behaviours are observed in the non-relativistic case for this problem. The saturation level will arguably be set by the strong feedback of large magnetic scales, but longer, and possible even more accurate simulations are needed to have a definitive answer.

\subsection{Spectra}

In addition to volume-integrated quantities,
whenever there is turbulence it is illustrative to compute also the radially-averaged spectrum \cite{durran17,paper1}. For a given field $f$ defined in a periodic box of side $L$, we use common {\tt python} functions to calculate its discrete fast Fourier transform $\hat{f}(\vec{k}) = \Sigma_{\vec{x}} f(\vec{x}) e^{-i \vec{k}\cdot\vec{x}}$, where the sum is performed over the $N^3$ spatial points equally spaced in each direction, with $k_j = n~\Delta k$, where $\Delta k = \frac{2\pi}{L}$ and $n\in [0,N/2]$ is an integer.

We consider the radial coordinates of the Fourier space, describing it with $\Delta k$-wide radial bins also centered on $k_r=\{n~\Delta k\}$. Then, we calculate the spectra $\mathcal{E}(k)$ as averages $<\cdot>_{k_r}$ over the annular bins of the power density per unit of radial wavenumber in 3D:

\begin{eqnarray}
&& \mathcal{E}_k(k_r) = \frac{L^3 4\pi}{(2\pi)^3N^6} <k^2|\widehat{\sqrt{\rho}\vec{v}}|^2(\vec{k})>_{k_r} \nonumber\\
&& \mathcal{E}_m(k_r) = \frac{L^3 4\pi }{(2\pi)^3N^6}<k^2|\hat{\vec{B}}|^2(\vec{k})>_{k_r} ~\label{eq:spectra}
\end{eqnarray}
where $k^2 = k_x^2+k_y^2+k_z^2$ and the normalization arises from the Parseval identity.
%. The normalization comes from the Parseval identity, so that $\int_0^{N/2} \mathcal{E}(k_r) dk_r = \int_V f^2(\vec{x}) dV$ where $f^2 = \rho v^2/2,B^2/2$ are the local energy densities in the real space. 
%Note that this identity, expressed in terms of the integral in $k_r$, strictly holds if the fluid is isotropic, so that the average over the annular bin does not lose statistically relevant information (for instance, a strong dependence on power on the direction of $\vec{k}$). 
For further technical considerations about normalizations, caveats (e.g. the systematic noise introduced by the conversion to radial coordinates in the Fourier space) and possible corrective factors, we refer to a recent dedicated paper \cite{durran17}. The analysis of our simulations with a large number of points have been parallelized by using the python package \cite{mortensen16} due to the large memory required.

The obtained spectra for $t=\{2,10,20\}$ are displayed in Fig.~\ref{fig:kh3d_spectra}. First of all, note that all spectra show a change of slope at $k \sim $ few times $2\pi/\Delta$, due to numerical dissipation. This depends on the scheme, and it is a feature that already appeared in other relativistic hydrodynamical turbulence works employing finite-differences \cite{radice13}. A possible cure could be to use spectral methods, which are suitable for bounding box simulations, but not for a complex astrophysical scenario. The change of slope means that the dynamics of the smallest resolvable scale is partially numerically damped. In the literature, this is actually a known issue that go under the name of implicit LES, meaning that the numerical dissipation effectively acts as an (uncontrollable) SGS Smagorinsky-like model within the simulation.
	
With this caveat in mind, we can analyze the spectra, focusing mostly on the large and intermediate scales. As time goes on, there is a direct cascade transferring kinetic energy from large to small scales (i.e., low to high wavenumbers). Therefore, the kinetic energy spectra extends to larger wavenumbers (i.e., smaller scales) with a Kolmogorov's slope $\propto k^{-5/3}$, as the resolution is increased.
	
When one looks at the magnetic spectra, one can see that most energy tends to be stored at small scales, because these are where they are injected via dynamo process. The magnetic energy is spread also at larger scales (what is known as inverse cascade), so that we obtain a similar shape for the magnetic energy (roughly compatible with the analytical Kazantsev dependence $\propto k^{3/2}$ \cite{kazantsev68}). The more we rise the resolution, the more effective is this dynamo mechanism. Since the physical limit to the small scale is set by the viscosity (which in turns set the thickness of the shear layer), which is neglected here, we can understand the lack of numerical convergence, at least in the growth phase. This explains the evolution of the total magnetic energy seen above.

Ideally, the implementation of an effective SGS model should properly include the feedback of the small-scale dynamics on the magnetic energy distribution and its growth. That should provide a magnetic spectrum which, at the intermediate and large scales, should be similar to a case with a much higher resolution (see our non-relativistic results in \cite{paper1}).

%%%%%%%%%%%%%%%%%%%%%%%%%%%%%%%%%%%%%%%%%%%%%%%%%%%%%%%%%%%%%
\section{A-priori fitting}\label{sec:apriori}
%%%%%%%%%%%%%%%%%%%%%%%%%%%%%%%%%%%%%%%%%%%%%%%%%%%%%%%%%%%%%

\subsection{Methodology}\label{sec:sfs_fit}
The most important analysis of our simulations is the a-priori test of the gradient SGS model.
We run simulations with a certain grid step $\Delta = L/N$. Then, we consider a snapshot at a given time, and spatially filter all the evolved fields. The simplest recipe is to use a simple average groups of $S_f^3$ cells, where we define $S_f$ as the filter factor. This corresponds to apply a filter in the real space, with a box kernel of size $\Delta_f = S_f\Delta$, obtaining filtered fields evaluated over $N_f^3=(N/S_f)^3$ points. The information lost in the filtering process is the field variation contained between the scales represented by $N_f$ and $N$. The solution in these SFS can be quantitatively evaluated by the explicit formal definitions of $\overline{\tau}$ as defined above in Eq.~\eqref{residual-F}. Let us consider a simple case as an illustrative example, with a SFS tensor defined as $\overline{\tau}(\vec{x}_f) = \overline{f}(\vec{x}_f)\overline{g}(\vec{x}_f) - \overline{fg}(\vec{x}_f)$. We can evaluate it at each of the $N_f^3$ positions of the filtered mesh $\{\vec{x}_f\}$, as 
\begin{equation}
\overline{\tau}(\vec{x}_f) = \frac{1}{S_f^{3}}\left[\Sigma_i f_i(\vec{x}_i) \Sigma_i g_i(\vec{x}_i) - \Sigma_i f_i(\vec{x}_i) g_i(\vec{x}_i)\right]  \label{eq:tau_example}
\end{equation}
where $i$ indicate each of the $S_f^3$ discrete positions considered inside the cell centered in $\vec{x}_f$.
Note that this estimation is not an exact evaluation of the loss information, since, by construction, it can only include the range of scales $[\Delta , S_f\Delta]$. The information for scales $<\Delta$ cannot be evaluated.

Once built each component of each SFS tensor, one can consider a given SGS model $\tau$. A measurement of the linear correlation between the numerical data and the different models can be estimated with the Pearson correlation coefficient between the SFS and SGS quantities, 
\begin{equation}
{\cal P} = {\rm Corr}\{\overline{\tau}(\vec{x}),\tau(\vec{x})\}
\end{equation}
While the Pearson coefficient tests the functional form, one can also consider each SGS component with a pre-coefficient $C$ to be adjusted. Its best-fit value  can be calculated by the minimizing the L2-norm, $\Sigma\, [\overline{\tau}(\vec{x}_f) - C\tau(\vec{x}_f)]^{2}$, where the sum is performed over all the positions $\{\vec{x}_f\}$. The minimization gives simply:
\begin{equation}
C_{\rm best} = \frac{\Sigma\,\overline{\tau}(\vec{x}_f)\,\tau(\vec{x}_f) }{\Sigma\,\tau(\vec{x}_f)^2}
\end{equation}
This procedure can be repeated independently for each SGS component, for each tensor $\tau$. As we showed in our previous work \cite{paper1}, when one compares the performance of the gradient model with other SGS models available for the non-relativistic case, the Pearson correlation of the gradient model stands out, being always much closer to one than the others, and degrading to $\lesssim 0.5$ only for quite large filter sizes.
 %Due to the usually very large of degrees of freedom ($N_f^3$), the associated $p$-value is almost always very small, and will not be taken into account as a useful indicator. Moreover, all the tested SGS models somehow depend on the derivatives of the field, which tend to correlate with the larger SFS residuals. Therefore, a certain statistically significant correlation can be expected in most cases. However, we need to identify the best-fitting model, and, in this sense, we will use the Pearson coefficients in a comparative way. 
%Remember that a positive value of these terms implies a local dissipation of energy (from resolved scales to SFS/SGS), and a negative value means a transfer from resolved to SFS/SGS.

Due to these previous findings and the absence of consistent models to compare with in the relativistic case, we assess the gradient model by a-priori fits for different times, resolutions, and filter size. Below, we report our main findings, usually represented by averaging the different independent components of a given tensor. Differences between different components are statistically negligible, and can arise only temporarily during the initial phase of the development when fields are not homogeneous (see \cite{paper1} for more details).

\subsection{Results}

As a representative example, we consider the different resolutions filtered with $S_f=2$ (thus, averaging to $N_f = 512^3$). In Fig.~\ref{fig:kh3d_bestfit}, we show the correlation Pearson values and the best-fit coefficients between the main SFS tensors, eqs.~\eqref{eq:rmhd_d}-\eqref{eq:rmhd_b} and the corresponding SGS components tensors, eqs.~\eqref{eq:sgs_gradient}, as a function of time.
The Pearson coefficients are very close to one for all tensor components (${\cal P}\gtrsim 0.8$) at all times, indicating a very good correlation between the SFS and the SGS tensors even during the transitional development of the instability (where we have the minimum value of the correlation). Moreover, the best-fit coefficients are fairly constant in time and $\sim {\cal O} (1)$. This is the most important result, which confirms that the proposed model is actually fitting well for a variety of time (i.e., MHD configurations) and resolutions. As a comparison, other SGS models in the non-relativstic cases studied in \cite{paper1} showed little correlation, or even not at all. In particular, the Smagorinsky model, the only one tested so far in GRMHD \cite{radice17}, showed  a non-zero correlation (${\cal P}\lesssim 0.3$ at its best, for $S_f=2$, with a largely-varying best pre-coefficient) due to the ability to catch the transfer from resolved to SFS scales. Notice however that this is only a part of the non-linear dynamics involved in MHD turbulence.

The overall great performance of the gradient model are well known \cite{grete15,grete16,grete17,grete17b,grete17phd,kessar16,paper1}, due to the strong mathematical basis on which it relies. 
Let us stress that we have explored different initial conditions, finding mainly the same results. Another interesting result can be obtained by studying the simulation with $N=1024^3$ points with different filter size $\Delta_f$, as displayed in Fig.~\ref{fig:kh3d_filtersize}. As the filter size increases, there is more information lost by averaging in the cells. This leads to a degradation of the Pearson coefficient, although it is still above ${\cal P}\gtrsim 0.5$ even for $\Delta_f \geq 16$, indicating that the LES gradient terms are able to fit the functional form of the non-linear terms corresponding to a quite higher resolution. The variation over time of the best-fit coefficients are limited, which legitimates one to consider a constant pre-coefficient in the LES implementation, without any dynamical procedure (like in other non-relativistic models \cite{germano91}) for its estimation.

%\CP{Yo eliminaria este parrafo}
%Contrarily to the non-relativistic case, the intrinsic complexity of the proposed SGS terms does not allow for a deep understanding and assessment of each term involved. However, some insight can be gained by looking at the relative weight of the numerical values of each SFS/SGS term. This will depend on the specific problem and time: for instance, terms including the magnetic field gradients will acquire more importance with time, due to the overall dynamo mechanism effects. Some of the terms are likely to be negligible in a specific time for a specific problem, but, in general, we need to include all of them, since we cannot predict precisely the complex non-linear behavior.

One can also assess the correlations between the auxiliary double gradient terms given by eqs.~\eqref{tau_p}-\eqref{Tauv} and the corresponding SFS terms. In Fig.~\ref{fig:kh3d_bestfit_aux}, we show as an illustrative example the good agreement obtained by comparing the SGS term $(-\xi H_v)$ with $(\widetilde{v} - \bar{v})$ (left panel), and $(-\xi H_\Theta)$ with $(\widetilde{\Theta} - \overline{\Theta})$ (right). This is an important check, since the contribution to the overall SGS tensors of Fig.~\ref{fig:kh3d_bestfit} could be given by a dominant component, thus ``hiding'' the others. Our results show that, taken one by one, each testable term (i.e., the $\propto H$ terms) correlates well with the corresponding SFS value.

%%%%%%%%%%%%%%%%%%%%%%%%%%%%%%%%%%%%%%%%%%%%
\section{Conclusions}
\label{sec:conclusions}
%%%%%%%%%%%%%%%%%%%%%%%%%%%%%%%%%%%%%%%%%%%%

In this paper we have presented a formalism to study LES for a generic system of conservation equations, in which the evolved fields appear through non-linear (and possibly not analytical) relations in the fluxes. We have also extended the gradient SGS model for this generic system of conservation laws.

%We have verified that, by using this formalism, one can recover known results, already available in the literature, when applied to non-relativistic MHD equations. 
We have verified that, by applying this formalism to the non-relativistic MHD equations, one recovers known results available in the literature.
Furthermore, we have applied the formalism to the relativistic MHD system, extending for the very first time 
%(i) the LES to the context of relativistic MHD and (ii) 
the gradient SGS model to the relativistic case. %\FC{esto me confunde un poco, la segunda no implica la primera?} \DV{Si, quitaria el (i), que ademas no es cierto... Una LES es cualquier simulacion que no sea una DNS (que captura todas las escalas), o sea, que tenga una resolucion mas grande que la escala de disipacion, con lo cual mirando la definicion todas las simulaciones de merger son LES en realidad. Mirad si se ha quedado en otro lado eso de decirle LES o DNS a los que no lo son, porque varias veces hemos pensado que una LES es una simulacion con SGS, y no es asi, segun la definicion.}\CP{ya esta}

We have performed 3D numerical simulations of the KHI in a bounding box at different resolutions with high-order numerical schemes. Within these simulations we have been able to compare the residual SFS tensors with the gradient SGS tensors (i.e., {\em a priori} tests), showing a high correlation between these two quantities, as indicated by a Pearson number close to 1. Notably, the resulting best fit-parameters are also close to the expected value $C_{best} \approx 1$. Finally, we have seen that the Pearson value of these tensors is still significant (i.e., larger than 0.5) with a filter size up to $\Delta_f = 16 \Delta$, indicating that this approach can effectively account for a resolution between a factor of 4 and one order of magnitude times larger.

Reproducing small-scale dynamics by means of a validated LES can be seen as a computationally efficient way to solve the equations, equivalent to consider an effective higher resolution. When implemented in a LES, this approach can guarantee a considerable saving of computational time, by using a relatively low resolution, or if combined with high resolution, a feasible way to capture previously unaccessible small-scale dynamics. Quantifying the gain is not trivial, since it depends on the numerical scheme used (see \cite{paper1} for considerations about the intrinsic dissipation in commonly-used finite-difference schemes). Our previous results in the non-relativistic case and the a-priori assessment here presented, allow us to estimate a gain in the effective resolution of at least a factor 4, and possibly up to a factor 8.

Finally, let us mention that although the LES and gradient SGS model presented here are valid only for the special relativistic case since we have used the Minkowski metric, the extension to General Relativity should be straightforward. As a matter of fact, although Einstein equations are highly non-linear, they do not show a ``turbulent'' regime: the metric tends to be smooth and slowly varying, compared to the matter fields. %\FC{No entendi eso ultimo, es correcto?} \DV{Quiero decir que las ec de Einstein no llevan a turbulencia. He quitado una parte de la frase que no entendia y puesto algo mas claro.}\CP{ya esta} 
Therefore, the metric functions entering in the filtered MHD equations vary much more smoothly than the fields involved in the turbulent dynamics. Such extension, together with its application to binary neutron star simulations, will be presented in a forthcoming paper.

\bibliography{turbulence}

%\appendix

\end{document}